\documentclass[a4paper,11pt]{article}
\pdfoutput=1

\usepackage{jcappub} 

\usepackage[T1]{fontenc} 
\usepackage[english]{babel}
\usepackage{geometry,amsmath,amsfonts}
\usepackage{slashed}
\usepackage{epsfig}
\usepackage{latexsym}
\usepackage{graphicx}
\usepackage{epstopdf}
\usepackage{amssymb}
\usepackage{subfig}
\usepackage{color}
\usepackage{multirow}
\usepackage{bbding}
\usepackage{rotating}
\usepackage{ifthen}
\usepackage{epsfig}
\usepackage{hyperref}
\ifx\pdfoutput\undefined
\usepackage[dvips,bookmarks=false]{hyperref}	
\else
\usepackage{hyperref}	
\fi
\usepackage{hyperref}
\hypersetup{
    colorlinks,
    citecolor=black,
    filecolor=black,
    linkcolor=black,
    urlcolor=black
}

\newcommand{\non}[0]{\nonumber \\}
\newcommand{\bee}[0]{\begin{eqnarray}}
\newcommand{\eee}[0]{\end{eqnarray}}

\newcommand{\be}[0]{\begin{equation}}
\newcommand{\ee}[0]{\end{equation}}

\renewcommand{\tilde}{\widetilde}

\title{Dark Matter in the minimal Inverse Seesaw mechanism}
\author[a]{Asmaa Abada,}
\author[b]{Giorgio Arcadi}
\author[a,c]{and Michele Lucente}

\affiliation[a]{Laboratoire de Physique Th\'eorique,\\
Universit\'e de Paris-Sud 11, B\^at. 210, 91405 Orsay Cedex, France}
\affiliation[b]{Institute for theoretical physics, 
Georg-August University G\"ottingen,\\
Friedrich-Hund-Platz 1, G\"ottingen, D-37077 Germany}
\affiliation[c]{Scuola Internazionale Superiore di Studi Avanzati,\\
via Bonomea 265, 34136 Trieste, Italy}

\emailAdd{asmaa.abada@th.u-psud.fr}
\emailAdd{arcadi@theorie.physik.uni-goettingen.de}
\emailAdd{michele.lucente@th.u-psud.fr}

\abstract{We consider the possibility of simultaneously addressing the dark matter problem and neutrino mass generation in the 
minimal inverse seesaw realisation. The Standard Model is extended by two right-handed neutrinos and three sterile fermionic states, leading to three light active neutrino mass eigenstates, two pairs of (heavy) pseudo-Dirac mass  eigenstates and one (mostly) sterile state with mass around the keV, possibly providing a  dark matter candidate, and accounting for the recently observed and still unidentified monochromatic 3.5 keV line  in galaxy cluster spectra. \\
The conventional production mechanism through oscillation from active neutrinos can account only for $\sim 43\%$ of the observed relic density. This can be slightly increased to $\sim 48\%$ when including effects of entropy injection from the decay of light (with mass below 20 GeV) pseudo-Dirac neutrinos. The correct relic density can be achieved through freeze-in from the decay of heavy (above the Higgs mass) pseudo-Dirac neutrinos. 
This production is only effective for a limited range of masses, such that the decay occurs not too far from the electroweak phase transition.
We thus propose a simple extension of the inverse seesaw framework, with an extra scalar singlet coupling to both the Higgs and the sterile neutrinos, which allows to achieve the correct dark matter abundance in a broader region of the parameter space, in particular in the low mass region for the pseudo-Dirac neutrinos.}


\arxivnumber{1406.6556}

\begin{document}
\maketitle
\flushbottom

\section{Introduction}
The origin of neutrino masses and the nature of dark matter are two of
the most pressing open questions of particle and  astroparticle physics. 

The basic structure of a 3-flavour leptonic mixing matrix, as well as the existence of two neutrino oscillation frequencies ($\Delta m_{ij}^2$) - in turn suggesting that at least two neutrinos have non-vanishing masses -, are strongly supported by current neutrino data (for a review, see~\cite{GonzalezGarcia:2012sz}). However,  
some of the current experiments (reactor~\cite{reactor:I}, accelerator~\cite{Aguilar:2001ty,miniboone:I}  and Gallium~\cite{gallium:I}) point to the 
existence of extra fermionic gauge singlets with masses in the eV range.
This would  imply that instead of  the three-neutrino mixing scheme, one would have a $3+1$-neutrino (or $3 +$more) mixing schemes (see, for instance,~\cite{Kopp:2013vaa}).

Sterile neutrinos are an intriguing and popular solution for the dark matter problem as well~\cite{Dodelson:1993je, Abazajian:2001nj,nu_WDM}. 

In particular, sterile neutrinos with masses around the keV can be viable Warm Dark Matter (WDM) candidates. They can potentially solve, even if providing only a fraction of the total dark matter (DM) relic density  some tensions with structure formation observations~\cite{general_structure}. In addition, a sterile neutrino at this mass scale  could  in general decay into an ordinary neutrino and a photon which could be detected in cosmic rays. This last possibility has recently triggered a great interest in view of the indication, yet to be confirmed, of an unidentified photon line in galaxy cluster spectra at an energy $\sim 3.5$~keV~\cite{Bulbul:2014sua,Boyarsky:2014jta}.  

In order to account for massive neutrinos, one of the most economical possibilities is to embed a seesaw 
mechanism~\cite{seesaw:I, seesaw:II, seesaw:III} into the Standard Model (SM). 
Low-scale seesaw mechanisms~\cite{ISS:ref,low_SS}, in which sterile fermions with masses around the electroweak scale or even lower are added to the SM particle content, are very attractive scenarios.
In particular, this is the case of the Inverse Seesaw mechanism (ISS)~\cite{ISS:ref}, which offers the possibility to accommodate
the smallness of the active neutrino masses $m_\nu$ for a comparatively  low seesaw scale, but still with natural
Yukawa couplings. The new  (light) states can be
produced in collider and/or low-energy experiments, and their
contribution to physical processes can be sizeable in certain realisations.
The  ISS mass generation mechanism
requires the simultaneous addition of both  right-handed (RH) neutrinos 
and extra sterile fermions to the SM 
field content ($\#\nu_R\ne 0$ and $\#s\ne 0$).\footnote{In the case where $\# s=0$, one recovers the type I  seesaw realisation which could account for neutrino masses and mixings provided that the number of right-handed neutrinos is at least $\#\nu_R=2$.} 
Being gauge singlets, and since there is no direct evidence for their existence,  the number of additional fermionic 
singlets $\# \nu_R$ and $\# {s}$ is unknown (moreover the numbers $\# \nu_R$ and $\# s$  are not necessarily equal~\cite{Abada:2014vea}). 
However,  these scenarios are severely constrained: any (inverse) seesaw realisation must comply with
a number of bounds. In addition to accommodating neutrino data (masses and mixings), they
must fulfil  unitarity bounds, laboratory bounds, electroweak (EW) 
precision tests, LHC constraints, bounds from rare decays, as well as 
cosmological constraints. 

In~\cite{Abada:2014vea},  
it was shown that it is possible to construct several minimal distinct ISS scenarios that can 
reproduce the correct neutrino mass spectrum while fulfilling all phenomenological constraints. 
Based on a perturbative approach, this study also showed that  
the mass spectrum of these minimal ISS realisations is characterised by either 2 or 3 different mass scales, 
corresponding to the one of the light active neutrinos $m_\nu$, that
corresponding to the heavy states $M_R$, and  an intermediate scale $\sim \mu$ 
only relevant 
when $\#s > \# \nu_R$. This allows to identify  two truly minimal ISS realisations (at tree level):
the first one, denoted "(2,2) ISS" model, corresponds to the SM extended by 
two RH neutrinos and two sterile states. It leads to a 3-flavour mixing scheme and requires only two scales (the light neutrino masses, $m_\nu$ and the RH neutrino masses, $M_R$). Although considerably fine tuned, this ISS configuration still complies with all phenomenological constraints, and systematically leads to a normal hierarchy for the light neutrinos.  
The second, the "(2,3) ISS" realisation, corresponds to an extension of the SM by 
two RH neutrinos and three sterile states, and  allows to accommodate both hierarchies for the light neutrino spectrum (with the  inverse hierarchy  only marginally allowed), in a  3+1-mixing scheme. The mass of the lightest sterile neutrino can vary over a large interval: 
depending on its regime, the 
"(2,3) ISS" realisation can offer an explanation for the 
short baseline (reactor/accelerator) anomaly~\cite{reactor:I,Aguilar:2001ty,miniboone:I,gallium:I} (for a mass of the lightest sterile state around the eV), or provide a WDM candidate (for a mass of the lightest sterile state in the keV range). 

In this work, we investigate in detail this last possibility, conducting a thorough analysis of the relic abundance of the warm dark matter candidate, taking into account all available phenomenological, astrophysical and cosmological constraints. The conventional DM production mechanism, the so called Dodelson-Widrow mechanism~\cite{Dodelson:1993je}, results in a tension with observational constraints from DM Indirect Detection (ID) and structure formation, since it can only  account for at most~$\sim 50\%$ of the total DM abundance. A sizeable DM density can  nonetheless be achieved when one considers the decay of the heavy pseudo-Dirac neutrinos. However this possibility is realised only in a  restricted region of the parameter space. An extension of the model is thus needed in order to account for a viable DM in a broader portion of the parameter space.       

The paper is organised as follows:  in Section~\ref{Sec:Model}, after  the description of  the model - the  "(2,3) ISS" realisation -,  we address  the prospects of the lightest sterile state as a viable DM candidate, which are  stability,  indirect detection and the dark matter generation mechanism. In Section~\ref{Sec:Requirements}, we  consider all the relevant  different astrophysical and cosmological  constraints taking into account the effect of the heaviest sterile neutrinos (DM production from  decays of heavy sterile states or possible entropy injection effects from a scenario with lighter sterile neutrinos) accounting as well for the indication of the monochromatic 3.5 keV observed line.
Section~\ref{Sec:NISS} is devoted to an economical extension of the present model which succeeds in providing   the remaining $\sim 50\%$ of  the dark matter relic abundance in a larger region of the parameter space.
Our final remarks are given in
Section~\ref{Sec: Conclusions}.  The numerical details regarding the production and evolution of the sterile neutrinos can be found in 
the Appendix. 

\section{Description of the model}\label{Sec:Model}

\subsection{The "(2,3) ISS" framework}
The inverse seesaw mechanism can be embedded into the framework of  the SM by introducing a mass term for neutrinos of the form:
\begin{equation}
-\mathcal{L}=\frac{1}{2}n_L^T C M n_L+\mbox{h.c.}\ ,
\end{equation}
where $C \equiv i \gamma^2 \gamma^0$ is the charge conjugation matrix and $n_L \equiv {\left(\nu_{L,\alpha},\nu_{R,i}^c,s_j\right)}^T$\!.  Here $\nu_{L,\alpha}$, $\alpha=e,\mu,\tau$, denotes the three SM left-handed neutrino states, while $\nu_{R,i}^c$ ($i=1,  \#\nu_R$) and $s_j$ ($j=1, \# s$) are additional right-handed neutrinos and  fermionic sterile states, respectively. 
Since there is no direct evidence for their existence,  and 
being gauge singlets,  the number of the additional fermionic 
sterile states $\nu_{R,i}$ and ${s_j}$ is unknown. 
In this work we will follow a bottom-up approach, focussing on the ISS realisations with the minimal content of extra right-handed neutrinos  and sterile fermionic states providing a viable phenomenology. 
 It is nonetheless worth mentioning that  a realisation of the ISS with 3 RH neutrinos and 4 sterile states fulfilling all possible constrains has been recently found in the context of conformal EW symmetry breaking~\cite{Lindner:2014oea}. 
 
The  neutrino mass matrix $M$ has the form:
\begin{equation}\label{general-iss}
M \equiv
\left(
\begin{array}{ccc}
0 & d & 0 \\
d^T & m & n \\
0 & n^T & \mu
\end{array}
\right)\ ,
\end{equation}
where $d,m,n,\mu$ are complex matrices.\\
The Dirac mass matrix $d$ arises from the Yukawa couplings to the SM Higgs boson $\tilde{H} =i  \sigma^2 H$,\begin{equation}
Y_{\alpha i} \overline{\ell_L^\alpha}\tilde{H }\nu_R^i+\text{h.c., }\,\,\,\,\,\,\ell_L^\alpha=\left(
\begin{array}{c}
\nu_{L}^\alpha \\
e^\alpha_L
\end{array}
\right)\ ,
\end{equation}
\noindent which gives after Electroweak Symmetry Breaking (EWSB):
\begin{equation}\label{Dirac}
d_{\alpha i}=\frac{v}{\sqrt{2}}Y^*_{\alpha i}\ .
\end{equation}  
The matrices $m$ and $\mu$ represent the Majorana mass terms for, respectively, right-handed and sterile fermions. Assigning a leptonic charge to both $\nu_R$ and $s$ with lepton number $L=+1$~\cite{ISS:ref}, in order that the Dirac mass term $-d^* \overline{\nu_L}\nu_R$ is lepton number conserving, the terms $\nu^T_R C \nu_R$ and $s^T Cs$ violate lepton number by two units. Within this definition,  the entries of $m$ and $\mu$ can be made small, as required for accommodating ${\mathcal{O}}(\text{eV}$) masses for ordinary neutrinos through the ISS mechanism. This does not  conflict with naturalness since the lepton number is restored in the limit $m,\mu \rightarrow 0$. In the following we will always work under the assumption that the magnitude of the physical
parameters (entries of the above matrices) fulfil such a naturalness criterium
\be
|m|,|\mu|\ll |d|,|n|\ , 
\ee
which implies the condition  $\#s \geq \# \nu_R$ for the mechanism to be viable~\cite{Abada:2014vea}.

Finally, the matrix $n$ represents  the lepton number conserving interactions between right-handed and new sterile fermions. 

The physical neutrino states $\nu_{I,\ I=1,\dots,3+\# \nu_R+\#s}$, are obtained upon diagonalization of the mass matrix $M$ via the unitary leptonic mixing matrix $U$,
\bee
n_L^T M n_L = \nu^T M_{\text{diag}}\ \nu,&& \left\{ \begin{array}{rcl} M_{\text{diag}} &=& U^T M U,\\
n_L &=& U \nu,\\
\end{array}\right.
\eee
and feature the following mass pattern\footnote{In the case of only one field for each kind, the light neutrino mass writes (to second order in perturbations of $|m|,|\mu|\ll |d|,|n|$)
\begin{equation}\label{1genneutrinomass(0)}
{m_\nu^2}^{(2)} \,=\, \frac{|d|^4 |\mu |^2}{\left(|d|^2+|n|^2\right)^2} \, ,\nonumber
\end{equation}
which reduces to the usual inverse seesaw expression once one assumes $|d| \ll |n|$. 
The first order corrections to ${m_{1,2}^2}^{(0)}= |d|^2+|n|^2$ lift the degeneracy in the heavy pseudo-Dirac states: 
\be\label{1genneutrinomass(1,2)}
\begin{array}{cc}
 {m_1^2}^{(1)} = -\frac{\left|\mu ^* n^2+m |d|^2+m |n|^2\right|}{\sqrt{|d|^2+|n|^2}}\,, \nonumber
 & {m_2^2}^{(1)}= \frac{\left|\mu ^* n^2+m |d|^2+m |n|^2\right|}{\sqrt{|d|^2+|n|^2}}\,.\nonumber
\end{array}
\ee
The expressions for the corresponding mass eigenstates, as well as the expressions for masses and corresponding eigenstates in the case of more than one field for each kind, are discussed in the appendices of \cite{Abada:2014vea}.
}~\cite{Abada:2014vea}:
\begin{itemize}
\item 3 light active states with masses of the form\begin{equation}\label{inv.ss}
m_\nu\approx \mathcal{O}(\mu) \frac{k^2}{1+k^2}\,, \,\,\,\,\,\,k\simeq\frac{\mathcal{O}(d)}{\mathcal{O}(n)}\ .
\end{equation}
This set must contain at least three different masses, in agreement with the two oscillation mass frequencies (the solar and the atmospheric ones). 
\item $ \# s- \# \nu_R$ light sterile states (present  only if $ \# s> \# \nu_R$) with masses $\mathcal{O}(\mu)$.
\item $\# \nu_R$ pairs of pseudo-Dirac heavy neutrinos with masses $\mathcal{O}\left(n\right)+\mathcal{O}\left(d\right)$.
\end{itemize}

In order to be phenomenologically viable, the matrix $M$ associated to a given ISS realisation must exhibit, upon diagonalization, three light, i.e. $\lesssim \mathcal{O}(\mbox{eV})$, active eigenstates with mass differences in agreement with oscillation data and a mixing pattern compatible with the experimental determination of the PMNS matrix.
 Although the minimal ISS realisation satisfying these requirements is the "(2,2) ISS", corresponding to  2 right-handed and 2 additional sterile fermions, a detailed numerical study performed in~\cite{Abada:2014vea} has  shown that complying with all constraints requires an important fine-tuning. 

On the contrary, a very good ``fit'' is provided by the "(2,3) ISS" realisation (2 right-handed and~3~additional sterile fermions). In this last case an additional intermediate state (with mass  $m_4=m_s=\mathcal{O}(\mu)$) appears in the mass spectrum.  
Remarkably, and in order to comply with all constraints from neutrino oscillation and laboratory experiments,  
the coupling of this new state  to the active neutrinos must be highly suppressed, thus leading to a dominantly  sterile state, with a mass ranging from $\mathcal{O}(\text{eV})$ to several tens of keV.\footnote{Light sterile neutrinos, i.e. with masses ranging between the eV and keV scale also appear in the so called Minimal Radiative Inverse See-Saw~\cite{Dev:2012bd}.} As consequence of its very weak interactions, the lifetime of the lightest sterile neutrino largely exceeds the lifetime of the Universe and  it can  thus play a relevant r\^ole in cosmology. 

In this work we will focus on the possibility that this sterile neutrino accounts, at least partially, for the Dark Matter component of the Universe, identifying the viable regions of the parameter space with respect to DM phenomenology of the "(2,3) ISS" model. 

We nonetheless mention that, in the lightest mass region (eV scale), and although not a viable DM candidate, the lightest sterile neutrino could potentially accommodate a $3+1$ mixing scheme explaining the (anti)-neutrino anomalies in short baseline, Gallium and reactor experiments~\cite{reactor:I,Aguilar:2001ty,miniboone:I,gallium:I}.

We also point out that the heaviest sterile states might be involved in a  broad variety of particle physics processes and have then to comply with several laboratory bounds and electroweak precision tests (these bounds have been analysed in~\cite{Abada:2014vea} for the "(2,2) ISS" and "(2,3) ISS" realisations). On recent times the possibility of production of heavy neutrinos at collider has been as well considered. The most peculiar signatures of the ISS scenario are, as a consequence of the large Yukawa couplings of the right-handed neutrinos, additional decay channels of the Higgs boson into a heavy and an ordinary neutrino, if kinematically allowed, or into three SM fermions through an off-shell neutrino. These decay modes can be searched both directly, in particular the ones with leptonic final states~\cite{BhupalDev:2012zg,higgs_nu}, and indirectly, in global fits of the Higgs data, by looking at deviations from the SM prediction in the branching ratios of the observed channels~\cite{BhupalDev:2012zg}. 
Direct searches of decay channels of the Higgs provide bounds on the Yukawa couplings of the pseudo-Dirac neutrinos with masses ranging from approximately 60 GeV (at lower masses possible signals do not pass current analysis cuts employed by experimental collaborations) to 200 GeV which can be as strong as $\sim 10^{-2}$ while global analysis of Higgs data provide a limit, for the same mass range, as strong as $\sim 3 \times 10^{-3}$ but can be effective in a broader mass range.
Alternatively heavy sterile neutrinos can be looked in dilepton~\cite{Kersten:2007vk} or dilepton+dijet processes~\cite{Atre:2009rg}, which are sensitive to their coupling to the W boson, that is related to the mixing between the active and the sterile neutrinos and thus provide bounds on the the elements of the mixing matrix $U$.
In the low mass region, namely $\lesssim \mathcal{O}(\mbox{GeV})$, heavy neutrinos can be detected in decays of mesons~\cite{Atre:2009rg,mesons_nu}. In this work we consider "(2,3) ISS" realisations satisfying the above experimental constraints. We remark that a sensitive improvement of these constraints in the low mass region is expected from the recently proposed SHIP~\cite{Bonivento:2013jag}. 

\subsection{Light sterile neutrino as Dark Matter}

Before the  analysis
we will briefly summarise the main issues that should be addressed in order for the lightest sterile neutrino to be a viable dark matter candidate.

\smallskip
\noindent {\bf Stability and Indirect Detection}:\\
The most basic requirement for a DM candidate is its stability (at least on cosmological scales).
All the extra neutrinos of the ISS model have a non zero mixing with ordinary matter. As a consequence, the lightest one is not totally stable  and can decay into an active neutrino and a photon $\gamma$. On the other hand, as already pointed out, its very small mixing makes the decay rate negligible with respect to cosmological scales. Nonetheless, a residual population of particles can decay at present times producing the characteristic signature of a monochromatic line in X-rays. This kind of signature is within reach of satellite detectors like CHANDRA and XMN which have put strong limits on the couplings between sterile and active neutrinos (due to  the lack of detection of this kind of signal). Recently,  the existence of an unidentified line in the combined spectrum of a  large set of X-ray galactic clusters has been reported~\cite{Bulbul:2014sua} and independently, in the combined observation of the Perseus Cluster and the M31 Galaxy~\cite{Boyarsky:2014jta}. These  observations can be compatible with the decay of a sterile neutrino with a mass of approximately 7 keV. 
 Confirmation of the latter result requires further observation, and most probably, higher resolution detectors like the forthcoming \emph{Astro-H}. 
 As we will show in the analysis, the "(2,3) ISS" model can account for this intriguing possibility; however, we will only impose that the sterile neutrino  lifetime does not exceed current observational limits. 
   
\smallskip
\noindent{\bf DM generation mechanism}:\\
The second issue to address is to provide a DM generation mechanism accounting for the experimental value of its abundance. In the pioneering work by Dodelson-Widrow (DW)~\cite{Dodelson:1993je}, it has been shown that the DM abundance can be achieved through active-sterile neutrino transitions.\footnote{The popular WIMP mechanism cannot be effective in our case  since sterile neutrinos could not exist in thermal equilibrium in the Early Universe due to  their suppressed interactions with ordinary matter.} This kind of production is always present provided that there is a non-vanishing  mixing between active and sterile neutrinos; as a consequence,  it is possible to constrain the latter as function of the neutrino mass by  imposing that the DM relic abundance does not exceed the observed value. The "(2,3) ISS" framework allows for an additional production mechanism, consisting in the decay of the heavy pseudo-Dirac states.
We will  discuss this point at a subsequent stage.    
 
\smallskip
\noindent{\bf Limits from structure formation}:\\
 Sterile neutrinos in the mass range relevant for the "(2,3) ISS" model are typically classified as warm dark matter. This class of candidates is subject to  strong constraints from structure formation, which typically translate into lower bounds on the DM mass.
We notice however, that the warm nature of the DM is actually related to the production mechanism determining the DM distribution function. Sterile neutrinos - with masses at the keV scale - produced by the DW mechanism can be considered as WDM; this may not be the case for other production mechanisms.

In the next section we will investigate whether the "(2,3) ISS" can provide a viable DM candidate.

\section{Dark matter production in the "(2,3) ISS"}\label{Sec:Requirements}
In this section we address the impact of the combination of three kinds of requirements on the DM properties on the "(2,3) ISS" parameter space.
The results presented below rely on the following hypothesis: a standard cosmological history is assumed with the exception of possible effects induced by the decays of heavy neutrinos; only the interactions and particle content of the "(2,3) ISS" extension of the SM are assumed.

Regarding DM production we will  not strictly impose that the relic abundance reproduces the observed  relic abundance, $\Omega_{\rm DM} h^2 \approx 0.12$~\cite{Ade:2013zuv},  but rather determine the maximal allowed DM fraction $f_{\rm WDM}$ within the framework of the "(2,3) ISS" parameter space.

The main production mechanism for DM is the DW, which is present as long as mixing with ordinary matter is switched on. In addition,  the DM could also be produced by the decays of the pseudo-Dirac neutrinos. However, a sizeable contribution can only be  obtained  if at least one of the pseudo-Dirac states lies in the mass range $130$ GeV - 1 TeV. Moreover, 
the pseudo-Dirac states can also have an indirect impact on the DM phenomenology since, under suitable conditions, they can release entropy at their decay, diluting the DM produced by active-sterile oscillations, as well as relaxing the bounds from structure formation. As will be shown below, this effect is also restricted to a  limited mass range for the pseudo-Dirac neutrinos.

For the  sake of simplicity and clarity, we  first discuss the case in which the heavy pseudo-Dirac neutrinos can be regarded as decoupled, and discuss at a second stage their impact on DM phenomenology. 

\subsection{Dark matter constraints without heavy neutrino decays}\label{sec:light_sterile}

We proceed to present the constraints from dark matter on the "(2,3) ISS" model, always  under the hypothesis that heavy neutrinos do not influence DM phenomenology. 

Regarding the relic density, for masses of the lightest-sterile neutrino with mass $m_s>0.1$~keV, we use the results\footnote{Notice that in~\cite{Asaka:2006nq},    the parametrisation $|M_D|_{\alpha 1} \equiv \theta_{\alpha s} m_s$ was used, while   in our work we use  $|U_{\alpha s}| \simeq \theta_{\alpha s}$ for  small mixing angles. } of~\cite{Asaka:2006nq}:
\begin{equation}\label{eq:Omega_s}
\Omega_{\rm DM} h^2 =1.1\times  10^7 \sum_\alpha  C_\alpha(m_s) \left|U_{\alpha s}\right|^2 \left(\frac{m_s}{\text{ keV}}\right)^2, \,\,\,\,\alpha=e,\mu,\tau\ .
\end{equation}
$C_\alpha$ are  active flavour-dependent coefficients\footnote{For DM masses of the order of 1 - 10~keV, the production peaks at temperatures of $\sim$150 MeV, 
corresponding  to the QCD phase transition in the primordial plasma. As a consequence,  
the numerical computation of the $C_\alpha$ coefficients is affected by uncertainties related to the determination of the rates of hadronic scatterings, and to the QCD equation of state.} which can be numerically computed by solving suitable Boltzmann equations. 
In the case of a sterile neutrino with mass $m_s<0.1$ keV, we have instead used the simpler expression~\cite{Abazajian:2001nj}:
\begin{equation}\label{eq:Omega_s_s}
\Omega_{\rm DM} h^2 = 0.3 \left(\frac{\sin^2 2 \theta}{10^{-10}}\right) \left(\frac{m_s}{100 \text{ keV}}\right)^2,
\end{equation}
where  $\sin^2 2 \theta = 4 \sum_{\alpha=e,\mu,\tau} |U_{\alpha s}|^2$, with  $|U_{\alpha s}|$  being the active-sterile leptonic mixing matrix element.
  We have then computed the DM relic density using Eqs.~(\ref{eq:Omega_s},\ref{eq:Omega_s_s})  for a set of "(2,3) ISS" configurations  satisfying data from neutrino oscillation experiment  and laboratory constraints. We have  imposed $f_{\rm WDM}=\Omega_{\rm DM}/\Omega_{\rm DM}^{\rm Planck} \leq 1$ thus obtaining constraints for  $m_s$ and $U_{\alpha s}$.  

The configurations with DM relic density not exceeding the experimental determination have been confronted with the limits coming from structure formation. 
There are several strategies to determine the impact of WDM on structure formation, leading to different constraints; in fact most of these constraints assume that the total DM component is accounted by WDM produced through the DW mechanism.  Notice that these constraints can  be relaxed when this hypothesis does not hold and we will address this point in a forthcoming section.

In the following, and when possible, we will thus reformulate the bounds from structure formation in terms of the quantity $f_{\rm WDM}$ which represents the amount of DM produced from active-sterile oscillation.\footnote{The results presented are in fact approximative estimates. A proper formulation would require detailed numerical studies, beyond the scope of this work.}   

The most solid bounds come from the analysis of the phase-space distribution of astrophysical objects. The WDM  free-streaming scale is of the order of the typical size of galaxies; as a consequence,  the formation of DM halos, as well as that of the associated galaxies is deeply influenced by the DM distribution function. According to this idea, it is possible to obtain  robust limits on the DM mass by requiring that the maximum of the dark matter distribution function inferred by  observation, the so called coarse grained phase space density, does not exceed the one of the fine-grained  density, which is theoretically determined and dependent on the specific DM candidate. 
Using this method, an absolute lower bound on the DM mass of around 0.3~keV, dubbed Tremaine-Gunn (TG) bound~\cite{Tremaine:1979we} was obtained by comparing the DM distribution from the observation of Dwarf Spheroidal Galaxies (Dphs) with the fine-grained distribution of a Fermi-gas. A devoted study of sterile neutrinos produced by DW mechanism has been presented in~\cite{Boyarsky:2008ju}, where a lower mass bound of the order of 2~keV was obtained. This limit can be evaded assuming that the WDM candidate is a subdominant component, while the DM halos are mostly determined by an unknown cold dark matter component. 
The reformulation of the limits in this kind of scenarios requires a dedicated study (an example can be found in~\cite{Anderhalden:2012qt}). In this work we conservatively rescale the results of~\cite{Boyarsky:2008ju} under the assumption that the observed phase-space density is simply multiplied by a factor $f_{\rm WDM}$. 
Moreover we have considered as viable the points of the "(2,3) ISS" model with $m_s < 2$~keV, featuring a value $f_{\rm WDM} \lesssim 1\%$, which corresponds approximatively to the current experimental uncertainty in the determination of the DM relic density.

For masses above 2~keV another severe bound is obtained from the analysis of the Lyman-$\alpha$ forest data. From these it is possible to indirectly infer the spectrum of matter density fluctuations, which are in turn determined by the DM properties. 
The Lyman-$\alpha$ constraint  is strongly model dependent and the bounds are related to the WDM production mechanism,  and to which extent this mechanism contributes to the total DM abundance. In order to properly take into account the possibility of only a partial contribution of the sterile neutrinos to the total DM abundance, we have adopted the results presented in~\cite{Boyarsky:2008xj} where the Lyman-$\alpha$ data have been considered in the case in which sterile neutrinos WDM account for the total DM abundance, as well as in the case in which they contribute only to a fraction (the remaining contribution being  originated by a cold DM component). More precisely, we have considered the most stringent 95\% exclusion limit\footnote{The limit considered actually relies on  data sets which are not  up-to-date. A more recent analysis~\cite{Viel:2013fqw} has put forward a stronger limit in the case of a pure WDM scenario, and thus the limits are  underestimated. As it will be clear in the following, the final picture is not  affected by this.}, expressed in terms of $(m_s, f_{\rm WDM})$, and translated it into an exclusion limit on the parameters of our model\footnote{Notice that the Lyman-$\alpha$ method is reliable for DM masses above 5~keV. For lower values there are very strong uncertainties and it is not possible to obtain solid bounds. In~\cite{Boyarsky:2008xj} it is argued  that the limit on $f_{\rm WDM}$ should not significantly change at lower masses with respect to the one obtained for neutrinos of 5~keV mass.}, namely the mass $m_s$ of the sterile neutrino and its effective mixing angle  with active neutrinos $\theta_s$.
We finally remark that WDM can be constrained also through other observations, as the number of observed satellites of the Milky way~\cite{DM_sat,Boyarsky:2012rt}, giving a lower bound on the DM mass of approximately 8.8~keV. 
This last kind of limits however strictly relies on the assumption that the whole dark matter abundance is totally originated by a WDM candidate produced through the Dodelson-Widrow mechanism and cannot straightforwardly be reformulated in case of a  deviation from this hypothesis;  thus we have not been considered these limits in our study. 

The inverse seesaw realisations passing the structure formation constraints have to be confronted to the limits from the X-ray searches, as reported in, for instance~\cite{Boyarsky:2012rt}. The corresponding constraints are  again given in the plane $(m_s,\theta_s)$, and can be schematically expressed by\footnote{Notice that the exclusion limit from X-rays is actually the combination of the outcome of different experiments and the dependence on the dark matter mass deviates in some regions from the one provided above. We have taken this effect into account in our analysis.}:
\begin{equation}
\label{eq:approx_X}
f_{\rm WDM} \sin^2 2\theta \lesssim  10^{-5} {\left(\frac{m_s}{1 \text{ keV}}\right)}^{\!-5}\ ,
\end{equation}    
where~\cite{nu_decay}:
\be 
\label{eq:pal}
\sin^2 2 \theta = \frac{16}{9} \sum_{i=1}^3 \left| \sum_{\alpha=e,\mu,\tau} U_{\alpha, s} \ U_{\alpha,i}^* F(r_\alpha)\right|^2, 
\,\,\,\,F(r_\alpha) = - \frac{3}{2} + \frac{3}{4} r_\alpha, \ r_\alpha=\left(\frac{m_\alpha}{M_W}\right)^2\ ,
\ee
with $i$ running over the different active neutrino mass eigenstates (3 different final states in the decay are possible). In the above expression we have again accounted for the possibility that the sterile neutrino contributes only partially to the DM component by rescaling the limit with a factor $f_{\rm WDM}$.\footnote{Notice that, contrary to the case of bounds from structure formation, this scaling is strictly valid only if the additional components does not decay into photons and thus it will not be applied in the next sections.} 

\begin{figure}[thb]
\begin{center}
\subfloat{\includegraphics[width=7.6cm]{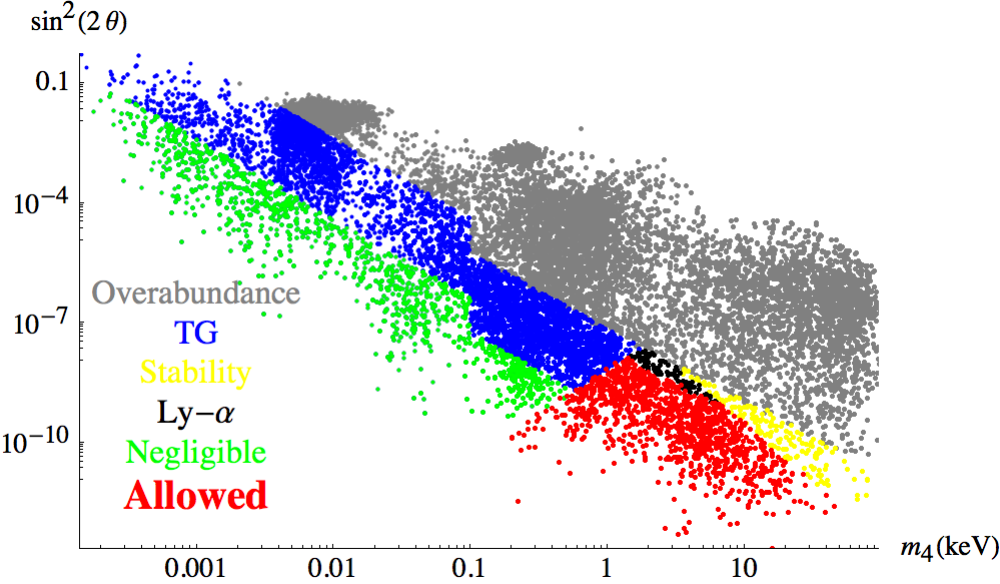}}\hspace*{0.5cm}
\subfloat{\includegraphics[width=7.6cm]{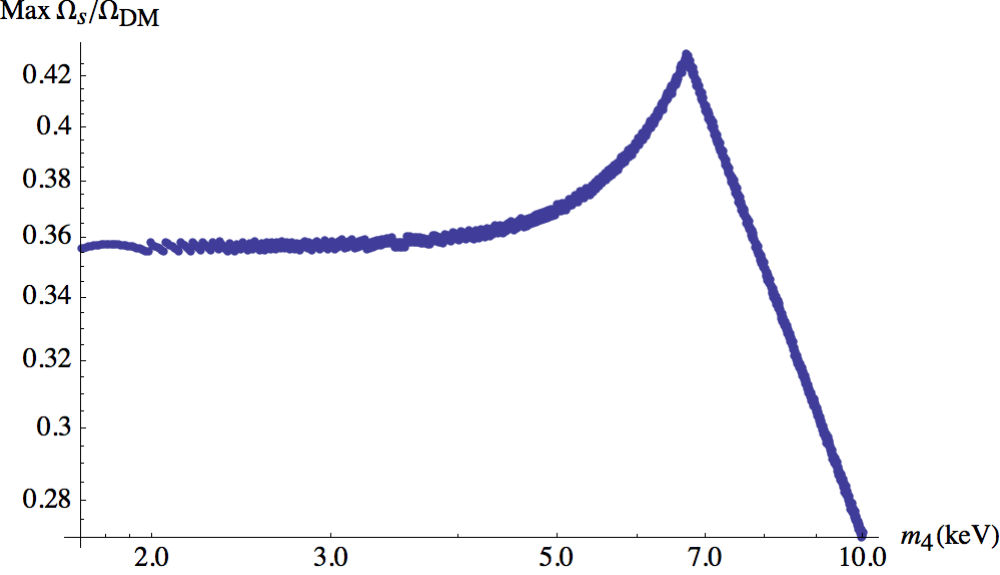}}
\end{center}
\caption{On the left panel, different regions of the lightest sterile neutrino parameter space $(m_4,\sin^2 2\theta$) identified by DM constraints. The grey region corresponds to a DM relic density exceeding the cosmological value. The blue, black and yellow regions are also excluded by phase space distribution, Lyman-$\alpha$ and X-ray searches constraints, respectively. The green region corresponds to configurations not excluded by cosmology but in which the lightest sterile neutrino contributes with a negligible amount to the DM relic density. Finally, the red region corresponds to the "(2,3)ISS"  configurations fulfilling  all the cosmological constraints, and for which the contribution to the dark matter relic density from the light sterile neutrino is sizeable. On the right panel, maximal value of $f_{\rm WDM}$ allowed by cosmological constraints as a function of the mass of the lightest sterile neutrino.}
\label{fig:summary_no_heavy}
\end{figure}

The result of the combination of the three kinds of constraints  applied in our analysis, namely dark matter relic density, structure formation and indirect detection, is reported in Figure~\ref{fig:summary_no_heavy}. As can be seen,  the requirement of a correct DM relic density has a very strong  impact, excluding a very large portion of the parameter space (grey region) at the highest values of the active-sterile mixing angles. Phase space density constraints rule-out most of   the configurations with mass of the lightest sterile neutrino below $\sim 2 \text{ keV}$ (blue region), a part a  narrow strip (green region) corresponding to $f_{\rm WDM} < 1 \%$. In this last region, and although not ruled out, the "(2,3) ISS" model cannot solve the Dark Matter puzzle, at least in its minimal realisation. In the large mass region, namely above 2~keV, a further exclusion comes form Lyman-$\alpha$ and indirect detection bounds (respectively black and yellow region) reducing the allowed active-sterile mixing. 
A sizeable contribution to the DM relic density can be thus achieved in a small localised region (in red)  of the parameter space, corresponding to masses of the lightest sterile neutrino in the range 2 - 50~keV and for active-sterile mixing angles $10^{-8} \lesssim \sin^2 2 \theta \lesssim 10^{-11}$.      
We show in the right panel of Figure~\ref{fig:summary_no_heavy} the maximal value of $f_{\rm WDM}$ allowed by the cosmological constraints as function of the DM mass. As  can be seen, the lightest sterile neutrino can only partially account for the DM component of the Universe with $f_{\rm WDM} \sim 0.43$ in the most favourable case. The maximal allowed DM fraction increases for the lowest values of the mass until a maximum at around 7~keV, after which it displays a  sharp decrease. This behaviour can be explained as follows: at lower masses, the Lyman-$\alpha$ bounds are the most effective and  become weaker as the mass of the sterile neutrino increases, thus allowing for larger $f_{\rm WDM}$. At the same time, the bounds from X-ray sources become stronger (since higher masses imply higher decay rates) thus reducing the allowed DM fraction as the mass increases.    

Notice that the above analysis is valid within the assumption that the production of the lightest sterile neutrino occurs in the  absence of a lepton asymmetry. Indeed, as firstly shown in~\cite{Shi:1998km}, the production of sterile neutrinos can be resonantly enhanced (as opposed to the conventional DW production usually called  non-resonant) in presence of a non-zero lepton asymmetry. In this case the correct dark matter abundance is achieved for much smaller active-sterile mixing angles, thus evading the limits from dark matter indirect detection; in addition, the resonant production alters the DM distribution function with respect to a non-resonant production, rendering  it  ``colder'' and thus compatible with Lyman-$\alpha$ constraints~\cite{Boyarsky:2008mt}. 

Interestingly a lepton asymmetry can be generated in frameworks featuring keV scale sterile neutrinos accompanied by heavier right-handed neutrinos. The entries of the active-sterile mixing matrix can in general be complex, and give rise to  CP-violating phases;  as a consequence, a lepton asymmetry can be generated by oscillation processes of the heavy neutrinos. In particular, it has been shown that a pair of quasi-degenerate right-handed neutrinos with masses of the order of a few GeV can generate a  lepton asymmetry before the EW phase transition (which is converted to the current baryon asymmetry of the Universe)  and then  at much later times, the lepton asymmetry needed to provide the correct relic density for a keV scale sterile neutrino~\cite{Shaposhnikov:2008pf,nuMSM,Canetti:2012kh}. 
The "(2,3) ISS" model also features pairs of quasi-degenerate heavy neutrinos which can be of the correct order of mass. However, the lepton asymmetry needed to ensure the correct DM relic density, compatible with the bounds discussed, requires an extreme degeneracy in  the heavy neutrino spectrum, of the order of the atmospheric mass differences. Such an extreme degeneracy is not achievable for the ISS model since  the predicted degeneracy of the pair of heavy neutrinos is of ${\mathcal{O}}(\mu)$, corresponding to around 1~keV for the cases under consideration.\footnote{Notice that a mass degeneracy of ${\mathcal{O}}(\text{keV})$ is still feasible for baryogenesis through oscillations of the heavy right-handed neutrinos.} A sizeable lepton asymmetry can be, however, generated by oscillation of not-degenerate neutrinos in the so-called flavoured leptogenesis~\cite{Akhmedov:1998qx} where individual lepton asymmetries in the different flavours are generated due to oscillations but the total lepton number is not necessarily violated. This mechanism has been, indeed, proven to be successful in explaining baryogenesis via leptogenesis thanks to sphaleron interactions~\cite{Canetti:2012kh,flavoured}, provided that there are at least three neutrinos contributing to the generation of the lepton asymmetry, and might be also efficient in generating the correct lepton asymmetry in order to have a resonantly enhanced DM production. This scenario is particularly promising in the "(2,3) ISS" model since it features four pseudo-Dirac neutrinos, potentially contributing to the generation of a lepton asymmetry. A quantitative investigation is however beyond the scope of the present work and is left for a future study. 

\subsection{Impact of the heavy pseudo-Dirac states}

The picture presented above can be altered in some regions of the parameter space due to the presence of the heavy neutrinos. Indeed, contrary to the DM candidate, they can exist in sizeable abundances in the Early Universe owing to their efficient Yukawa interactions, and influence the DM phenomenology through their decays.
There are two possibilities. The first one is  direct DM production from decays mediated by Yukawa couplings.
The branching ratio of these processes is small when compared to that of other decay channels into SM states, since it is suppressed by the  small active-sterile mixing angle, an efficient DM production can nevertheless be achieved through the so called freeze-in mechanisms if the pseudo-Dirac neutrinos are heavier than the Higgs boson.  
Significantly lighter pseudo-Dirac neutrinos, namely with masses below~$\sim 20\,\mbox{ GeV}$, can instead indirectly  affect DM phenomenology. Indeed,  they can be  sufficiently long-lived such that they can dominate the energy density of the Universe,  injecting entropy at the moment of their decay. 
We will discuss separately these two possibilities in the next subsections.  

\subsubsection{Effects of entropy injection}   

The conventional limits on sterile neutrino DM can be in principle evaded in presence of an entropy production following the decay of massive states dominating the energy density of the Universe~\cite{Scherrer:1984fd}. A phase of entropy injection dilutes the abundance of the species already present in the thermal bath and, in particular, the one of DM if such an entropy injection occurs after its production. In addition, the DM momentum distribution gets redshifted - resembling a ``colder'' DM candidate - and   suffering weaker limits from Lyman-$\alpha$. This phase of entropy injection can be triggered in the "(2,3) ISS" model by the decay of the heavy pseudo-Dirac neutrinos if the following two conditions are realised:  
 Firstly, at least some of the heavy sterile neutrinos should be sufficiently abundant to dominate the energy budget of the Universe. Secondly they must decay after the peak of dark matter production, but before the onset of  Big Bang Nucleosynthesis (BBN). These two requirements will identify a  limited region of the parameter space outside which the results of the previous subsection strictly apply.  

All the massive eigenstates have Yukawa interactions with ordinary matter described by an effective coupling $Y_{\rm eff}$ which is defined by:
\begin{equation}
Y_{\alpha \beta} \,\overline{\ell_L^\alpha} \,\tilde{H }\, \nu_R^\beta = Y_{\alpha \beta} \, \overline{\ell_L}^\alpha \,\tilde{H }\, U_{\beta i} \nu_i
=Y_{\rm eff}^{\alpha i}\  \overline{\ell_L^\alpha} \,\tilde{H }\,  \nu_i\ .
\end{equation}
These interactions are mostly efficient at high temperature when scattering processes involving the Higgs boson and top quarks are energetically allowed;  in addition, they maintain the pseudo-Dirac neutrinos in thermal equilibrium until temperatures of the order of $\sim 100 \text{ GeV}$, provided that $Y_{\rm eff}^2 \gtrsim 10^{-14}$~\cite{Akhmedov:1998qx}. If this condition is satisfied, an equilibrium abundance of heavy pseudo-Dirac neutrinos existed at the early stages of the evolution of the Universe.   

The Yukawa interactions become less efficient as the temperature decreases. At low temperature  the transition processes from the light active neutrinos become important. For a given neutrino state, the rate of the transition processes reaches a maximum at around~\cite{Asaka:2006ek}:
\begin{equation}
\label{eq:T_N_D}
T_{\text{max},I} \simeq 130 {\left(\frac{m_I}{1\text{ keV}}\right)}^{\frac{1}{3}} \text{MeV}\ .
\end{equation}

The transition rate of each neutrino at the temperature $T_{\text{max},I}$ exceeds the Hubble expansion rate $H$ if~\cite{Asaka:2006ek}:
\begin{equation}
\label{eq:Heavy_equilibrium}
\theta > 5 \times 10^{-4} \left(\frac{1\text{ keV}}{m_I}\right)^{1/2}\ , 
\end{equation} 
and thus, if this condition is satisfied, the corresponding pseudo-Dirac neutrinos are in thermal equilibrium in an interval of temperatures around $T_{\text{max},I}$. 

Notice that the picture depicted above assumes that the production of sterile neutrinos from oscillations of the active ones is energetically allowed; as a consequence it is valid only for neutrino masses lower than $T_{\rm max}$:
\bee\label{eq:condition}
T_{\text{max},I} \simeq 130 {\left(\frac{m_I}{1\text{ keV}}\right)}^{\frac{1}{3}} \text{MeV} \geq m_{I} &\Rightarrow & m_I \leq m_{I,\text{max}} \approx  46.87 \text{ GeV}.
\eee
As will be made clear in the following, neutrinos heavier than $M_{I,\text{max}}$ have excessively large decay rates to affect DM production and  hence  will not be relevant in the subsequent analysis.

\begin{figure}[thb]
 \begin{center}
\subfloat{\includegraphics[width=7.6cm]{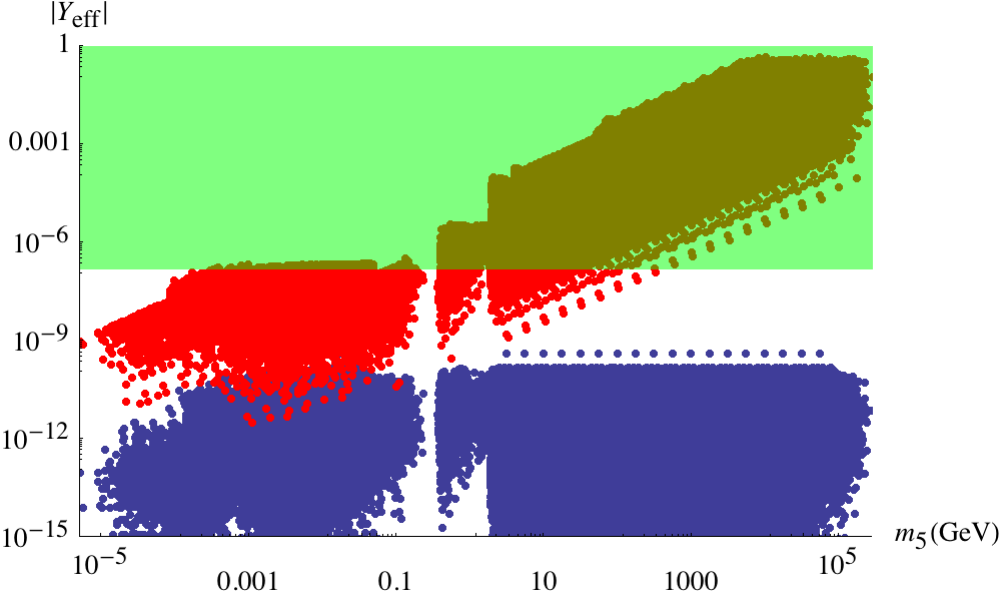}}\hspace*{0.6cm}
\subfloat{\includegraphics[width=7.6cm]{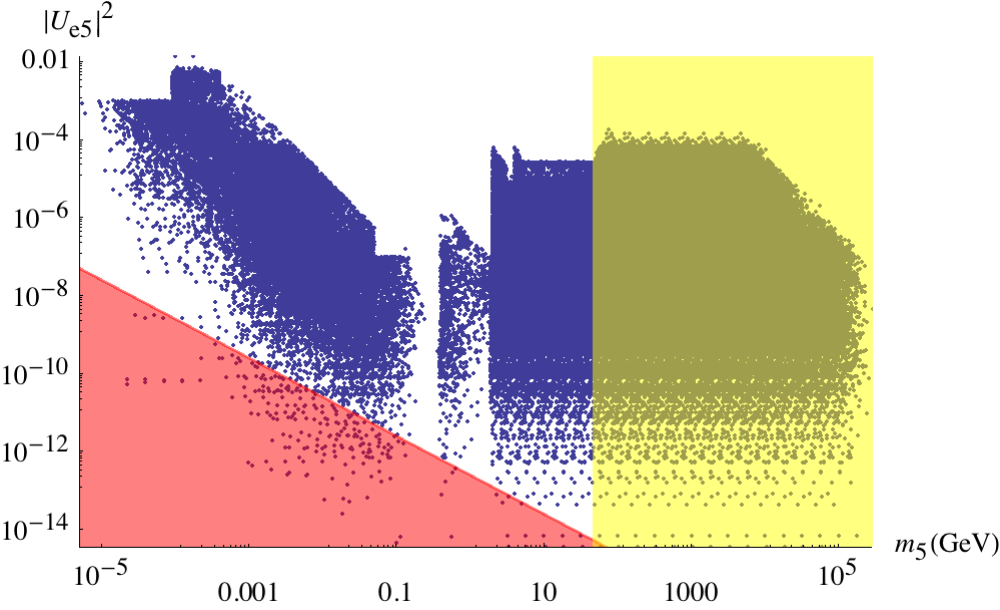}}
\end{center}
\caption{On the left panel: effective Yukawa couplings $Y_{\rm eff}$ for the neutrino DM candidate (blue~points) and of the lightest pseudo-Dirac particle (red points), as a function of the mass $m_5$.
The green region corresponds to   values $Y_{\rm eff}>\sqrt{2}\times 10^{-7}$, the limit above which the states are in thermal equilibrium. On the right panel: mixing of the electron neutrino with the lightest pseudo-Dirac state as a function of its mass. 
 The yellow region corresponds to the kinematically forbidden values of the sterile mass, see  Eq.~(\ref{eq:condition}).
 The red region denotes the solutions not in thermal equilibrium.}
\label{heavy_unbroken}
\end{figure}

In Figure~\ref{heavy_unbroken}, we present the typical behaviour of the pseudo-Dirac neutrinos in the regimes of high and low temperatures, which are dominated, respectively, by Yukawa interactions and active-sterile transitions. In the left panel of Figure~\ref{heavy_unbroken}, we display  the values of the mass and effective Yukawa couplings ($Y_{\rm eff}$) of the lightest pseudo-Dirac state (the other heavy states exhibit an analogous behaviour), corresponding to a set of "(2,3) ISS" realisations compatible with laboratory tests of neutrino physics (red points). The green region translates the equilibrium condition for the Yukawa interactions. In the larger  mass region, i.e. for masses significantly larger than 10 GeV, the value of the effective Yukawa coupling $Y_{\rm eff}$ is always above the equilibrium limit and can even be of order one for  higher values of the mass. In this region,  the pseudo-Dirac neutrinos can have  a WIMP-like behaviour and can be in thermal equilibrium until low temperatures. As already mentioned neutrinos in this mass range have impact on Higgs phenomenology at the LHC; we have compared the configurations of Figure~\ref{heavy_unbroken} with the limits presented e.g. in~\cite{BhupalDev:2012zg} and found they all result viable. 
 In  the intermediate mass region, i.e. for masses  between 1 and  a few tens of GeV,  equilibrium configurations are still present. However the values of $Y_{\rm eff}$ are lower with respect to the previous case and the decoupling of the pseudo-Dirac neutrinos depends on the oscillation processes at low temperatures. Configurations for which $Y_{\rm eff}$ is too small to ensure the existence of a thermal population of pseudo-Dirac neutrinos in the early Universe (they can be nonetheless created by oscillations at lower temperatures) are also present. This last kind of configurations are the only ones corresponding to  masses below 0.1~GeV.
We emphasise that the outcome discussed here  is a direct consequence of the "(2,3) ISS" mechanism which allows to generate the viable  active neutrino mass spectrum for pseudo-Dirac neutrinos with masses of the order of the EW scale, and for large values of their Yukawa couplings.
For comparison, we display in the same plot the distribution of values of the effective Yukawa couplings of the WDM candidate as a function of the mass of next-to-lightest sterile state $m_5$  (blue points). As can be seen, the corresponding solutions are always far from thermal equilibrium due to the suppressed mixing $U_{\nu_R,4}$.  

In the right panel of Figure~\ref{heavy_unbroken}, we display  the mixing (for small angles it is possible to approximate $\theta_{e 5} \simeq U_{e 5}$) of the lightest pseudo-Dirac state with the electron neutrino as a function of  $m_5$  for the ISS realisations compatible with laboratory limits.  The yellow region corresponds to the values of the sterile mass for which the DW production mechanism is kinematically forbidden  (see Eq.~(\ref{eq:condition})).  The red region denotes the solutions which are not in thermal equilibrium.

Combining the results obtained from the  two panels of Figure~\ref{heavy_unbroken}, we can conclude that all the considered realisations in the relevant mass interval satisfy the equilibrium conditions.  Consequently we can always assume the presence of an equilibrium population of the pseudo-Dirac states up to temperatures of the order of $T_{\text{max},I}$. We stress that $T_{\text{max},I}$ is not the actual decoupling temperature that has been instead determined  in for instance~\cite{Shaposhnikov:2008pf}  and more recently in~\cite{Canetti:2012kh}, and which turns out to be lower than $T_{\text{max}}$; however this affects only marginally our discussion.

As already pointed out, we will be interested in masses of the pseudo-Dirac neutrinos not exceeding 10 - 20~GeV. For  such a mass range we can safely assume that the neutrinos decouple when they are relativistic (see Eq.~(\ref{eq:T_N_D})) and that their decay occurs at a much later stage, when they become non-relativistic, as described in~\cite{Asaka:2006ek}.

In this setup, the pseudo-Dirac states can dominate the energy budget of the Universe if their energy density, which is defined by
\begin{equation}
\rho_N(T) \equiv \sum_{I=5,8} m_I n_I(T),\,\,\,\,\,\,n_I(T)=\frac{g_{*}(T)}{g_{*}(T_D)}{\left(\frac{T}{T_D}\right)}^3 n_I^{\rm eq}(T_D)=\frac{g_{*}(T)}{g_{*}(T_D)} \frac{3 \zeta(3)}{2 \pi^2} T^3\ ,
\end{equation} 
exceeds the radiation energy density $\rho_r=\frac{\pi^2}{30}g_{*}(T)T^4$, where $g_{*}(T)$ represents the number of relativistic degrees of freedom at the temperature $T$. 
Provided that the pseudo-Dirac neutrinos are sufficiently long-lived, this occurs at a temperature $\overline{T}$ given by:
\begin{equation}
\label{eq:Tbar}
\overline{T} \approx 6.4 \, \mbox{MeV} \left(\frac{m_5}{1 \text{GeV}}\right)\left(\frac{\sum_{I} m_I Y_I}{m_5 Y_5}\right)\ ,
\end{equation}
where we have taken $g_{*}(T_D)=86.25$ and $m_5$ is the mass of the lightest pseudo-Dirac neutrino. In this scenario the decay of the pseudo-Dirac neutrinos is accompanied by a sizeable amount of entropy; the conventional radiation dominated era restarts at the reheating temperature $T_{r,I}$~\cite{Scherrer:1984fd},  
and the abundance of the species present in the primordial thermal bath is diluted by a factor $S$, which is defined as the ratio of the entropy densities of the primordial plasma at temperatures immediately below and above the reheating one.

Notice that the above discussion corresponds to a simplified limit: in general the four pseudo-Dirac neutrinos have different masses and different lifetimes. 
In the "(2,3) ISS" model the pseudo-Dirac states appear as pairs with the mass splitting in each pair much smaller than the masses of the corresponding states. Identifying the  mass scale of each pair as  $m$ and $M$, with $m < M$, we can write, to a  good approximation\footnote{The discussion of this section, as well as the expressions here presented, are valid in the so called "instantaneous reheating approximation" which assumes that the entropy injection occurs at the reheating temperature. In fact the entropy release is a continuous process and the quantities $T_{r,M/m}$ and $S_{M,m}$ are not determined analytically but  extrapolated from the numerical solution of suitable Boltzmann equations~\cite{Arcadi:2011ev}. Moreover at high temperatures, namely $T \gtrsim m_I$, the decay rate of massive states into radiation is altered by effects from, for example, thermal masses or quantum statistical effects~\cite{not_inst_reheat} and the prediction for the reheating temperature might sensitively deviate from the prediction obtained in the instantaneous reheating approximation~\cite{Drewes:2014pfa}.      

In the setup under consideration we assume the pseudo-Dirac neutrinos decoupling at the temperature $T_{\rm max,I}$ defined in~(\ref{eq:T_N_D}). For the range of masses of pseudo-Dirac neutrinos for which the active-sterile transitions are effective $T_{\rm max,I} >m_I$ and increases while $m_I$ gets lower. In particular we have that $T_{\rm max}/m_I \sim 10$ for $m_I \sim 1\,\mbox{GeV}$. 

On the other hand, comparing the decay rates given in~(\ref{eq:Gammah}) and~(\ref{eq:Gammaz}), it results that the decay temperatures of the pseudo-Dirac neutrinos are lower than the masses of the neutrinos themselves at least for $m_I \lesssim 10\,\mbox{GeV}$ but they can be even lower by considering $Y_{\rm eff} \lesssim 10^{-3}$. The most relevant impact from the decays of the pseudo-Dirac neutrinos is obtained for very low decay temperatures, for which it is reasonable to neglect thermal corrections. Our main results can be thus described by the instantaneous reheating approximation.}:
\begin{equation}
 S=  S_{ m}\,  S_{M}\ ,
 \label{entropytot}
\end{equation}  
where $S_{ m}$ and $S_{M}$ are the dilution factors associated to the decays of the two pairs of pseudo-Dirac states, occurring at the two reheating temperatures $T_{r,M}$ and $T_{r,m}$, given by:
\begin{align}
\label{eq:entropy_release}
& S_M= {\left[1+2.95  {\left(\frac{2 \pi^2}{45}g_{*}(T_{r,M})\right)}^{1/3}{\left(\frac{\sum_{\alpha} m_\alpha Y_\alpha}{M\ Y_M}\right)}^{1/3}\frac{{(M\ Y_M)}^{4/3}}{(\Gamma_M M_{\rm PL})^{2/3}}\right]}^{3/4}\ ,  \nonumber\\
& S_{ m}= {\left[1+2.95  {\left(\frac{2 \pi^2}{45}g_{*}(T_{r,m})\right)}^{1/3}2^{1/3}\frac{{\left(\frac{m\ Y_m}{S_{ M}}\right)}^{4/3}}{(\Gamma_m M_{\rm PL})^{2/3}}\right]}^{3/4}\ ,
\end{align}
where $\Gamma_{m,M}$ is the decay rate of the heavy neutrinos. Notice that in the last term of each of the above equations, the effects of the first entropy dilution  have been included in the abundance of the lightest pair of heavy neutrinos.

The DM phenomenology is affected only when the pseudo-Dirac neutrinos dominate the Universe and decay after DM production. For keV scale DM,  this translates into the requirement $T_{r,m} \lesssim 150$ MeV. On the other hand,  a very late reheating phase would alter the population of thermal active neutrinos, leading to modifications of some quantities such as the primordial Helium abundance~\cite{Kawasaki:2000en} and the effective number of neutrinos $N_{\rm eff}$, and producing effects in structure formation  as well. By combining BBN and CMB data\footnote{There are also further cosmological constraints on heavy neutrinos derived using different approaches, see for instance~\cite{BBN:bound}.} it is possible to determine a solid bound $T_{r,m}>4$ MeV~\cite{Hannestad:2004px}. In addition, we have  considered  a (relaxed) limit of~$T_{r,m}>0.7$ MeV by  taking into account the possibility that this bound is evaded when the decaying state can produce ordinary neutrinos~\cite{Fuller:2011qy}. This choice is also motivated by the fact that after the decay of the neutrinos with mass $M$, the ratio $\rho_I/\rho_r$ between the energy densities of the remaining neutrinos and of the radiation is of order 3 - 5, implying that although subdominant, the radiation component is sizeable.  

The requirement $4 (0.7)\text{ MeV} \leq T_{r,m} \leq 150\text{ MeV}$ is satisfied only for a very restricted pseudo-Dirac neutrinos mass range. Indeed, sterile neutrinos can decay into SM particles through three-body processes mediated by the Higgs boson with a rate:
\begin{equation}
\label{eq:Gammah}
\Gamma_h= \frac{Y_{\rm eff}^2 \ m_I^5}{384 {\left(2 \pi\right)}^3 m_h^4} \sum_f y_f^2 \left(1-\frac{4 m_f^2}{m_I^2}\right)\ ,
\end{equation}
which implies an excessively  short lifetime for sterile neutrinos unless their masses are below (approximately) $\mathcal{O}(10\text{ GeV})$, in such a way that the decays into third generation quarks are kinematically forbidden and $Y_{\rm eff}$ can assume lower values. At these smaller masses, a sizeable contribution comes from $Z$ mediated processes with a rate:
\begin{equation}
\label{eq:Gammaz}
\Gamma_Z = \frac{G_F^2 m_I^5 \sin^2\theta_I}{192 \pi^3}\ , 
\end{equation}
where we have defined, for simplicity, effective mixing angles $\theta_I,I=m,M$ between the pseudo-Dirac neutrinos and the active ones. 

\begin{figure}[thb]
\begin{center}
\subfloat{\includegraphics[width=4.8cm]{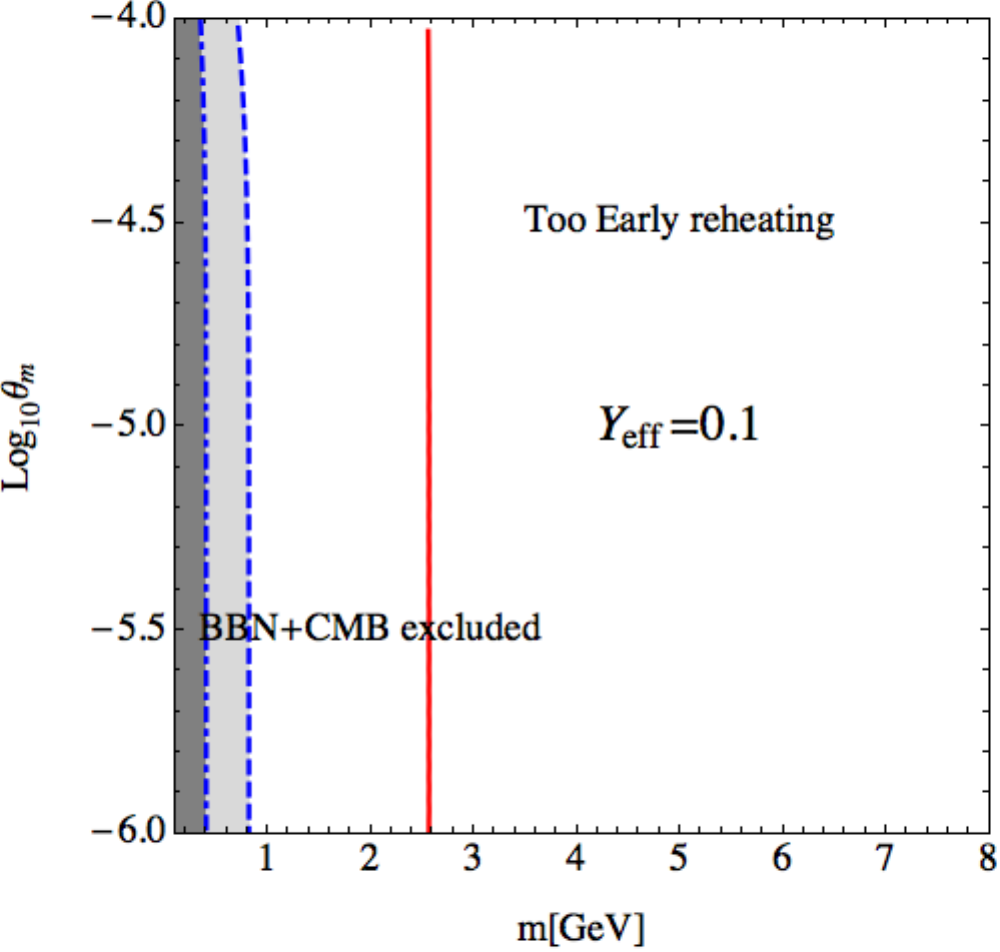}}\hspace*{0.4cm}
\subfloat{\includegraphics[width=4.8cm]{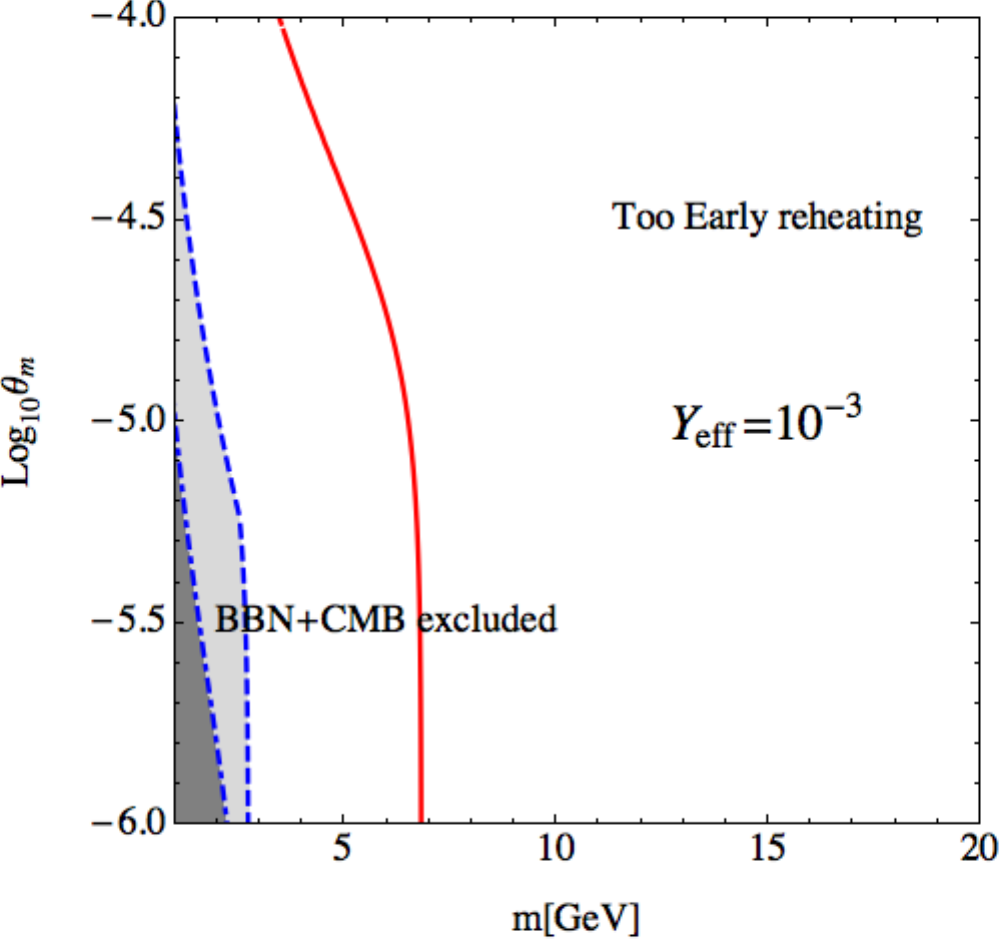}}\hspace*{0.4cm}
\subfloat{\includegraphics[width=4.8cm]{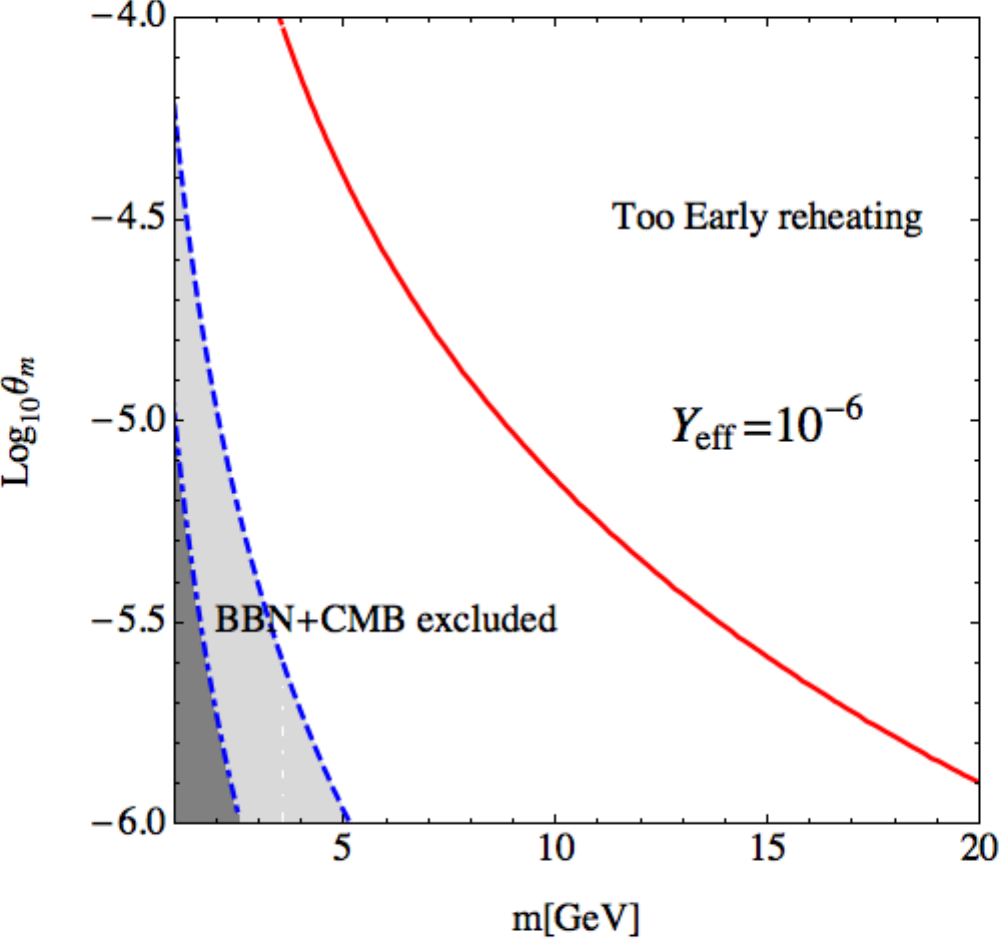}}
\end{center}
\caption{Region of the $(m,\theta_m)$ parameter space in which the decay of the pseudo-Dirac states can affect the DM phenomenology for three values of $Y_{\rm eff}$,  0.1, $10^{-3}$ and $10^{-6}$. The red lines in the panels represent $T_{r,m} = 150$ MeV. Above this line the reheating takes place before the DM production. The grey region below the dashed (dot-dashed) blue line is excluded by BBN/CMB combined constraints according to the limit $T_{r,m} > 4\ (0.7)$ MeV.}
\label{fig:Tr_limits}
\end{figure}

We have reported in Figure~\ref{fig:Tr_limits} the limit values of the lower reheating temperature $T_{\rm r, m}$ as a function of the mass scale $m$ and the effective mixing angle $\theta_m$, for three values of $Y_{\rm eff}$, namely 0.1, $10^{-3}$ and $10^{-6}$. The regions above the red curves correspond to an excessively large reheating temperature which does not affect DM production. 
The light-grey (dark-grey) regions below the blue curves represent values of the reheating temperature in conflict with the conservative (relaxed) cosmological limit of 4 (0.7) MeV. For ``natural'', i.e. $\mathcal{O}(1)$, values of the effective coupling, the decay rate of the heavy neutrinos is dominated by the Higgs channel and tends to be  too large except for  a narrow strip at masses of $1 - 2$ GeV. At lower values of $Y_{\rm eff}$ the size of the region corresponding to the interval 4 (0.7) - 150 MeV of reheating temperatures increases. The contours corresponding to the lower values of the reheating temperature are mildly affected by the values of $Y_{\rm eff}$ since,  at lower masses, the Higgs channel is suppressed by the Yukawa couplings of the first generation, in comparable amount with the $Z$ channel.
As can be seen from the left panel of  Figure~\ref{heavy_unbroken}, laboratory constraints favour values of $Y_{\rm eff} < 10^{-3}$ in the mass range $1 - 20$ GeV thus favouring  the possibility of an impact of the heavy neutrinos decays on the DM phenomenology,  in this region.

\begin{figure}[thb]
\begin{center}
\includegraphics[width=7.5cm]{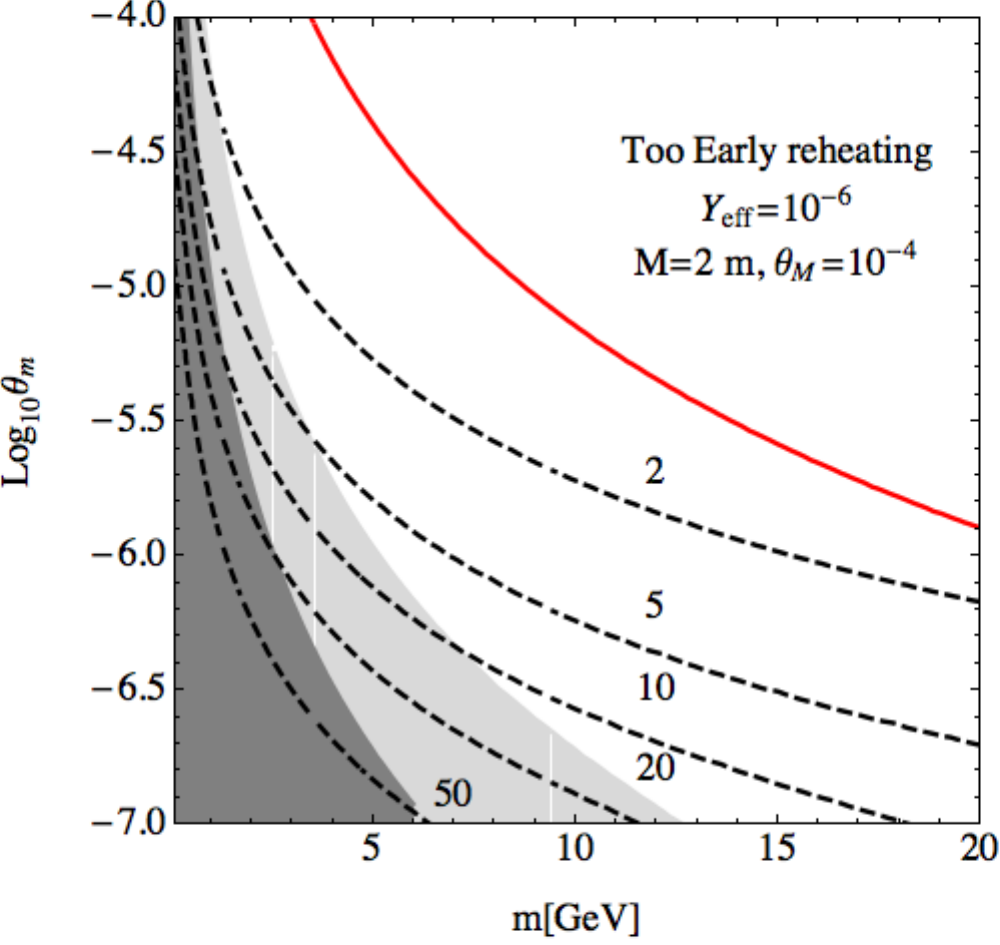}
\end{center}
\caption{Iso-curves of entropy injection in the plane $(m,\theta_m)$ with the remaining parameters $M$, $\theta_M$ and $Y_\text{eff}$ fixed according to the reported values. The grey (light grey) region corresponds to the lowest reheating temperature below 4 (0.7) MeV and is thus in tension with the cosmological bounds. In the region above the red curve the entropy injection takes place before the DM production.}
\label{fig:DeltaS}
\end{figure}

In Figure~\ref{fig:DeltaS} we have estimated the range of values of $S$  in the allowed parameter space, see Eq.~(\ref{entropytot}). In order to illustrate, we have chosen  to vary   $m$ and $ \theta_m$ fixing  $M$ to be $M=2\times m$ and $\theta_M=10^{-4}$, which corresponds to a  viable case for the "(2,3) ISS" mass spectrum; we have also  fixed the effective Yukawa coupling as $Y_\text{eff}=10^{-6}$ in order to maximise the phenomenologically relevant region of the parameter space.

As can be seen from Figure~\ref{fig:DeltaS},  we have only very moderate values of the entropy injection when the conservative lower limit of 4 MeV  is imposed on the reheating temperature; values of $S$ up to around 20 can be achieved once a weaker bound is considered. 

Having determined the range of variation of the entropy dilution within the "(2,3) ISS" parameter space,  we have reformulated the limits on the DM mass and mixing angle, as presented in the previous section, for the case where $S>1$.  
The limits from DM relic density can be straightforwardly determined  by simply rescaling it by a factor $1/S$. The limits from Lyman-$\alpha$ are more difficult to  address since this would require a different analysis (as for example~\cite{Boyarsky:2008xj,Boyarsky:2008mt}), which is computationally demanding and lies beyond the scope of this work. 
To a good approximation, one can assume a redshift factor of $S^{1/3}$~\cite{Asaka:2006ek} for the DM momentum distribution and translate it into a modified limit for the DM mass given by $m_{s,Ly\alpha}^S \geq m_{s,Ly\alpha} /S^{1/3}$, where $m_{s,Ly\alpha}$ is the lower limit on the DM mass for a given value $f_{\rm WDM}$ of the DM fraction, in the case where  $S=1$.   

The X-ray limits remain  unchanged with respect to the previous section since these rely on the DM lifetime. Nevertheless, the entropy injection leads to an indirect effect since a given pair $(m_s,\theta)$ now corresponds in general to a lower relic density.   

\begin{figure}[thb]
\begin{center}
\subfloat{\includegraphics[width=7.6cm]{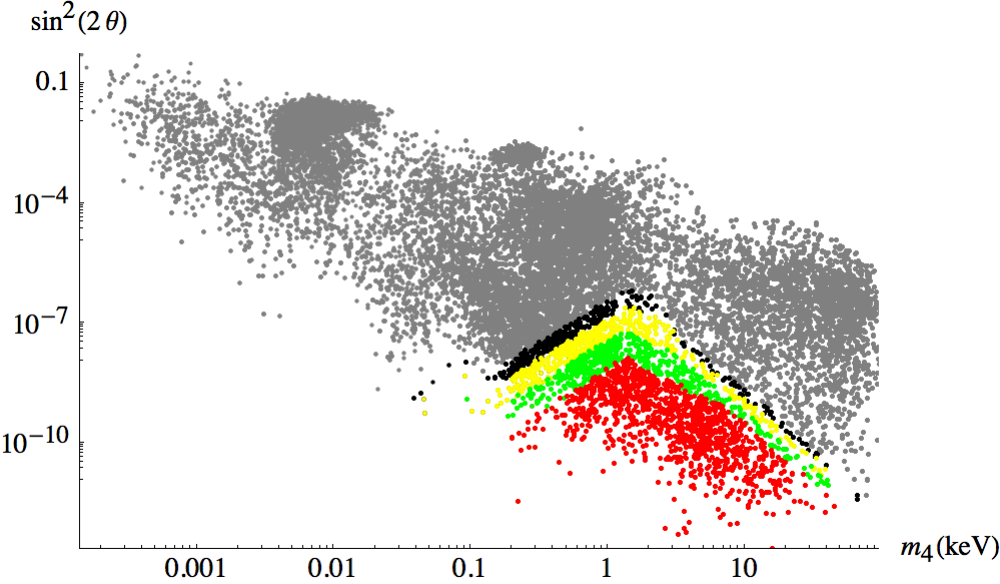}}\hspace*{0.4cm}
\subfloat{\includegraphics[width=7.6cm]{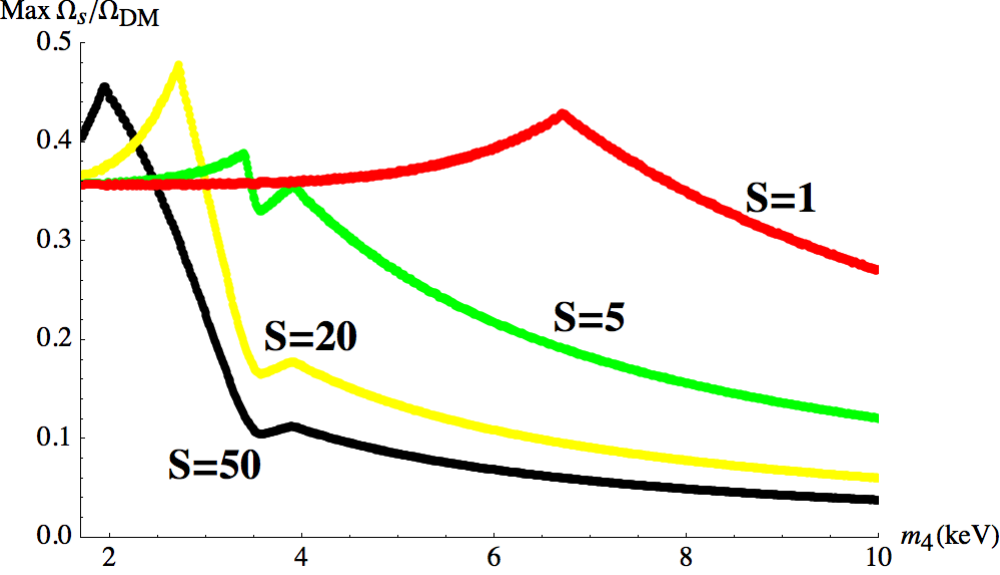}}
 \caption{\emph{Left panel}: Parameter space for the light sterile neutrino compatible with cosmological bounds in the hypothesis of an entropy injection, for values S=1 (red), S=5 (green), S=20 (yellow) and S=50 (black).  \emph{Right panel}: Maximal values of $f_{\rm WDM}$ allowed by the cosmological bounds as a function of the sterile neutrino mass in the hypothesis of an entropy injection $S$.}
\label{fig:entropy_abundance}
\end{center}
\end{figure}

The outcome of our analysis is summarised in Figure~\ref{fig:entropy_abundance}. 
In the left panel of Figure~\ref{fig:entropy_abundance},  we display the cosmologically favoured region for several values of  $S \leq 50$ as compared to the case $S=1$, represented by  red points. Values of $S$ larger than $\sim 20$ are not within reach in the framework of the present  model, but we have nonetheless extended our analysis up to these values in order to infer the maximal extension of the parameter space which could be achieved. The grey points are  excluded by DM constraints unless its relic density is negligible (see previous subsection). As one can see,  for $S>1$ we have a larger range of allowed values for the active-sterile mixing angle; interestingly, the augmentation of $S$ has a finite effect in enlarging the available parameter space.

On the right panel of Figure~\ref{fig:entropy_abundance}, we display  the maximal value of $f_{\rm WDM}$ for several values  of $S$ (we also display  for comparison the case corresponding to $S=1$). As one can see, there is only a marginal increase, namely from 0.43 to 0.48, of the maximal allowed value of $f_{\rm WDM}$. On the other hand, the maximal DM fraction is achieved for smaller  values of the  allowed DM mass, namely $\sim 2$~keV, as opposed to  values around $\sim7$~keV in the case where $S=1$. We finally notice that the maximal DM fraction is not an increasing function of $S$ but on the contrary, a maximum achieved at $S=20$ is followed by a sharp decrease. 
The reason of such  a behaviour is  mostly due to  the X-ray exclusion. Indeed, as already pointed out, any fixed value $f_{\rm WDM}$ imposes a condition on  $(m_s, \sin^2 2\theta)$ which is not sensitive to the mechanism accounting for the DM generation (more specifically, the value of $S$ in our case). 
Since the dark matter generation mechanism also depends on  $(m_s, \sin^2 2\theta)$, the interplay with the X-ray exclusion, as well as the effect of entropy injection,  favours larger mixing angles (thus maximising the  production of dark matter) and lower values of the mass (which in turn minimise the DM decay rate). 
Our analysis shows that  $f_{\rm WDM}=1$ could be achieved for  $S\lesssim 10$. This is not sufficient to relax the Lyman-$\alpha$ bound down to the value $m_s=2\text{ keV}$ because of the scaling of the latter limit as $S^{1/3}$. 

We finally point out that for very high values of $S$, sizeable values of $f_{\rm WDM}$ could be  achieved for very large mixing angles, already excluded by indirect dark matter  detection. This is at the origin of the saturation of the cosmologically favoured region observed in the left panel of 
Figure~\ref{fig:entropy_abundance}.

\subsubsection{Dark Matter Production from heavy neutrino decays and the 3.5 keV line}
As already mentioned, the pseudo-Dirac neutrinos can produce dark matter through their decays. These processes are mediated by  Yukawa interactions and the decay rate is proportional to $Y_{\rm eff} \sin\theta$, and thus suppressed with respect to the decay channels into only SM particles by the active-sterile mixing angle. A sizeable DM production can be nonetheless achieved through the so called freeze-in mechanism~\cite{FIMP}. It consists in the production of the DM while the heavy neutrinos are still in thermal equilibrium and, to be effective, requires that the rate of decay into DM is very suppressed, such that it results lower than the Hubble expansion rate. In our setup,  this condition can be expressed as: $Y_\text{eff} \sin\theta < 10^{-7}$. 

\begin{figure}[thb]
\begin{center}
\includegraphics[width=8.5 cm]{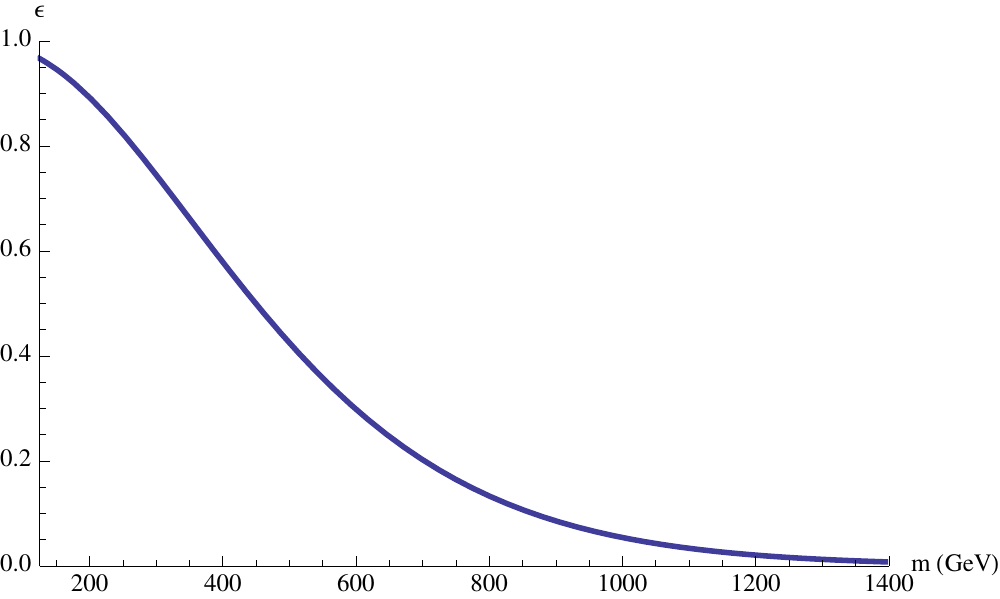}
\caption{Suppression factor in Eq.~(\ref{eq:heavy_fimp}) due to the electroweak symmetry restoration at high temperatures, as a function of the mass of the decaying particle.}
\label{fig:chi}
\end{center}
\end{figure}

The dark matter relic density depends on  the decay rate of the pseudo-Dirac neutrinos into DM as follows:
\begin{equation}
\label{eq:general_fimp}
\Omega_{\rm DM}h^2 \simeq \frac{1.07 \times 10^{27}}{g_{*}^{3/2}}\sum_I g_I \frac{m_{\rm s} \Gamma\left(N_I \rightarrow \mbox{DM}+\mbox{ anything}\right)}{m_I^2}\ ,
\end{equation}
where the sum runs over the pseudo-Dirac states and $g_I$ represents the number of internal degrees of freedom of each state.
For pseudo-Dirac neutrinos lighter than the Higgs boson, DM production occurs through three-body processes whose rate is too suppressed to generate a sizeable amount of DM.  
On the other hand, the above analytical expression  is not strictly applicable for heavier pseudo-Dirac neutrinos since the mixing angle $\theta$ depends on the vacuum expectation value (vev) of the Higgs boson and is thus zero above the EW phase transition temperature. To a good approximation, the correct DM relic density is determined by multiplying Eq.~(\ref{eq:general_fimp}) by  the function $\varepsilon(m_I)$ given by:
\begin{equation}
\varepsilon(m_I)=\frac{2}{3\pi}\int_{0}^{\infty} f(x_I) x_I^3 K_1(x_I) dx_I,\,\,\,\,\,x_I=\frac{m_I}{T}\, ,
\end{equation}  
with $f(x_I)$ describing the evolution of the Higgs vev $v(T)$ with the temperature and which can be in turn approximated, according the results presented in~\cite{D'Onofrio:2014kta}, by:
\begin{equation}
\frac{v(T)}{v(T=0)}\ =\ \left \{
\begin{array}{cc}
1 & T < T_{\rm EW} \\
8 - \frac{m_I}{20 x_I} & T_{\rm EW} \leq T \leq 160\text{ GeV} \\
0 & T > 160 \text{ GeV}\ 
\end{array}
\right. \ ,
\end{equation} 
where $T_{\rm EW}\approx 140$ GeV is the temperature associated to the EW phase transition. 
As shown in Figure~\ref{fig:chi},  the function $\varepsilon(m_I)$ sharply decreases with the mass of the pseudo-Dirac neutrino since  most of the FIMP (Feebly Interacting Massive Particle) production occurs around the mass of the decaying particle. As a consequence, we can have sizeable production of DM only for masses of the decaying particles not too much above the scale of the electroweak  phase transition while DM production is  negligible for masses of the pseudo-Dirac neutrinos above the TeV scale.  
Using the expression of the rate associated to the process $N_I \rightarrow h \ +  \text{ DM}$:
\begin{equation}\label{eq:NhDM}
\Gamma\left(N_I \rightarrow h \ + \text{ DM}\right)=\frac{m_I}{16 \pi} Y^2_{\rm eff,I} \sin^2 \theta \left(1-\frac{m_h^2}{m_I^2}\right)\ , 
\end{equation}
the DM relic density is given by:
\begin{equation}
\label{eq:heavy_fimp}
\Omega_{\rm DM} h^2 \approx 2.16 \times 10^{-1} {\left(\frac{\sin\theta}{10^{-6}}\right)}^2 {\left(\frac{m_{\rm s}}{1 \text{ keV}}\right)} \sum_I g_I {\left(\frac{Y_{\rm eff,I}}{0.1}\right)}^2 {\left(\frac{m_I}{1 \mbox{TeV}}\right)}^{-1} \left(1-\frac{m_h^2}{m_I^2}\right)\varepsilon\left(m_I\right).
\end{equation}
It is then clear that the correct DM relic density can be achieved with a suitable choice of the parameters. It is worth noticing  that this production mechanism is  complementary to the DW one, which is always active provided that there is a nonzero active-sterile mixing. 

\begin{figure}[thb]
\begin{center}
\includegraphics[width=9cm]{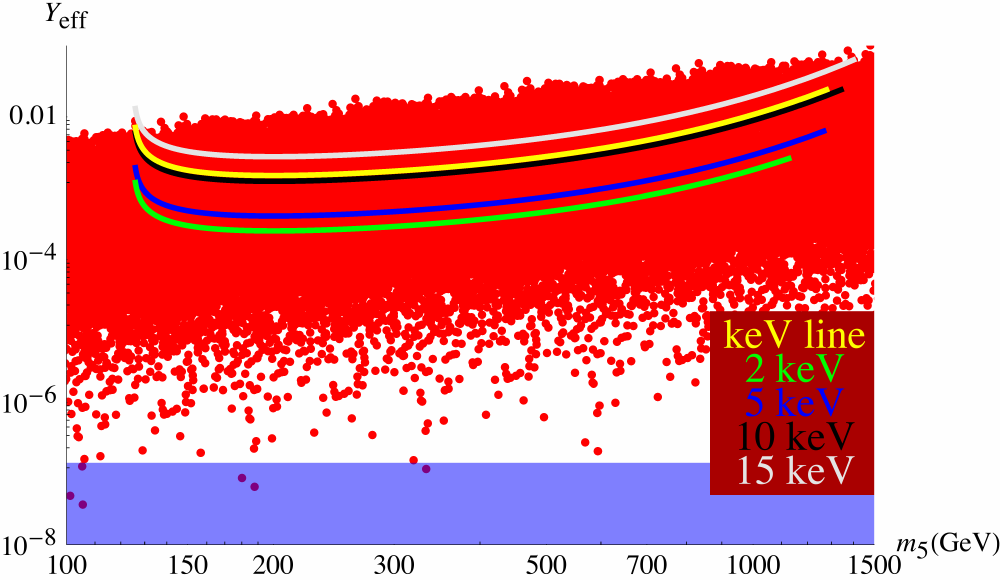}
\caption{
Viable configurations (continuous lines) for the heavy pseudo-Dirac masses $m_5$ and the corresponding effective Yukawa couplings $Y_\text{eff}$ accounting for the observed dark matter abundance of light sterile neutrinos via the freeze-in production mechanism, with masses and mixings for the sterile neutrinos compatible with cosmological bounds.
The red points denote different realisations of the "(2,3) ISS" model. In the blue region the production is not effective since the pseudo-Dirac states are out of thermal equilibrium. The lines end when the condition $Y_\text{eff}\  \sin\theta<10^{-7}$ is violated.
The yellow line accounts for the still unidentified monochromatic 3.5~keV line in galaxy cluster spectra~\cite{Bulbul:2014sua,Boyarsky:2014jta}.
}
\label{fig:3.5kev}
\end{center}
\end{figure}

We have reported in Figure~\ref{fig:3.5kev} the (observed) value $\Omega_{\rm DM} h^2=0.12$ of the DM abundance, assuming for simplicity the same mass  $m_5$  and effective Yukawa couplings $Y_\text{eff}$ for the 4 heavy pseudo-Dirac states, 
for  different values of the DM mass and considering the maximal value of $\sin\theta$ allowed by cosmological constraints - including thus the corresponding contribution from DW production  mechanism-.
The displayed red points correspond to configurations of the "(2,3) ISS" model in agreement with all  laboratory constraints. Those configurations corresponding to pseudo-Dirac states far from thermal equilibrium, and  thus not accounting  for a freeze-in production mechanism, are delimited by a blue region. 
The shape of the lines can be understood as follows: for pseudo-Dirac masses comparable with the Higgs one,  the kinematical suppression in Eq.~(\ref{eq:heavy_fimp}) is significant, requiring sizeable Yukawas; for $m_I \gtrsim 200\mbox{ GeV}$ the dependence on $m_I$ is weaker, and the curve reaches a plateau, while for $m_I \gtrsim 500\mbox{ GeV}$ the suppression due to the function $\varepsilon\left(m_I\right)$ becomes significant requiring larger Yukawas, eventually violating the freeze-in condition $Y_\text{eff}\  \sin\theta<10^{-7}$ for $m_I \gtrsim 1.2\text{ TeV}$.

The requirement of light  sub-eV  active neutrino masses together with $\mu \approx$ keV and $M_R \approx v$, implies values for the Yukawa couplings in the appropriate range to accounting for the observed DM abundance  ($m_\nu \approx \mu Y^2 v^2/M_R^2$, see Eq.~(\ref{inv.ss})). We emphasize here that this is not the case for a type-I seesaw realisation  since in this case the relation $m_\nu\approx Y^2 v^2/M_R < 1$ eV implies $Y\lesssim10^{-6}$ if $M_R \approx v$, and thus the contribution from the freeze-in process is not sufficient to account for the total DM abundance.

Among the lines displayed in  Figure~\ref{fig:3.5kev}, we have highlighted in yellow  the one corresponding to the following  DM mass and mixing angle, 
\bee\label{eq:3.5kev}
m_s \simeq 7.1 \text{ keV},&\phantom{XXX}&
\sin^2 2 \theta \approx 7 \cdot 10^{-11},
\eee
which can account for the monochromatic 3.5~keV line  observed in the combined spectrum of several astrophysical objects~\cite{Bulbul:2014sua,Boyarsky:2014jta}. 

The results presented in Figure~\ref{fig:3.5kev} do not take into account the possible constraints  from structure formation.  As will be made clear in the next section, the  limits discussed  above should be sensitively relaxed since the DM produced through the freeze-in mechanism has a ``colder'' distribution with respect to the DW mechanism. A reformulation of the corresponding  limits  is beyond the scope of this work, especially in the case in which the DM production receives sizeable contributions both from DW and decay of the pseudo-Dirac neutrinos. 
We argue nonetheless  that the parameters accounting for the keV line can be compatible with bounds from structure formation since for this choice (of parameters),  the DM abundance is entirely determined by the decay of the heavy neutrinos (the DW contribution for that value of the mixing angle is less than 4\%) and the corresponding distribution function is  ``colder'' with respect to the one of a resonantly produced DM, which results compatible with the observational limits~\cite{Abazajian:2014gza}.

To summarise the results obtained and discussed in this section, one can  state that  in the  absence of effects from the heavy pseudo-Dirac neutrinos, the "(2,3) ISS" model can, in the most favourable case, account for  to approximatively the $\sim 43 \%$ of the total DM density for a mass of approximatively 7~keV. This percentage slightly increases up to $48\%$, for a DM mass of around 2~keV, once accounting for an entropy dilution factor of 5 - 20 which can be possible for masses of the pseudo-Dirac neutrinos of 3 - 10~GeV. The total DM component can be accounted for only in the region $m_h < m_I< 1.4\text{ TeV}$, when the DM can be produced through the freeze-in mechanism, although the compatibility with structure formation should be still  addressed. 
In order to also reproduce the correct relic density  for masses of the sterile neutrinos below  the Higgs boson  mass, it is necessary to extend the particle field content  of the model;  for this purpose, we will propose  in the following section a minimal  extension of the "(2,3) ISS" model.  

\section{Dark Matter Production in minimal extension of the "(2,3) ISS" model}
\label{Sec:NISS}
In order  to achieve the correct dark matter  relic density in  the pseudo-Dirac states low mass regime,  we consider a minimal extension of the "(2,3) ISS" model. This  consists in the introduction of a scalar field $\Sigma$, singlet under the  SM gauge group, interacting only with the sterile fermionic states and the Higgs boson. There are of course several other possibilities, see for instance~\cite{singlet}. In this minimal extension, the part of the  Lagrangian where the new singlet scalar field is involved reads:  
\begin{equation}
\label{eq:Sigma_lagrangian}
\mathcal{L}=\frac{1}{2}\partial_\mu \Sigma \partial^\mu \Sigma-\frac{h_{\alpha \alpha}}{2}\Sigma\overline{s^c}_{\alpha}s_{\alpha}+V(H,\Sigma).
\end{equation} 
We consider that the field $\Sigma$ has a non-vanishing vev $\langle \Sigma \rangle$ that would be at the origin of  the Majorana mass coupling  $\mu$ which can  thus be expressed as:
\begin{equation}\label{mumu}
\mu \simeq 1 \text{ keV} \left(\frac{\langle \Sigma \rangle}{100 \text{ GeV}}\right) \left(\frac{h_{\alpha\alpha}}{10^{-8}}\right)\ .
\end{equation} 
For simplicity we will limit the scalar potential to the following terms (see e.g.~\cite{Petraki:2007gq} for a more general discussion):
\begin{equation}
V(H,\Sigma)=-\mu^2_{H} |H|^2-\frac{1}{2} \mu_{\rm \Sigma}^2 \Sigma^2 +2\lambda_{\rm H\Sigma}|H|^2 \Sigma^2\ .
\end{equation}
Following a pure phenomenological approach, we will consider values of the portal coupling $\lambda_{H\Sigma}$ from order of $10^{-2}$, corresponding to limits from effects on the Higgs width~\cite{Kamenik:2012hn}, down to very low values, i.e. ${\cal{O}}\left(10^{-8}\ \text{or}\  10^{-9}\right)$ (see for instance ~\cite{Merle:2013wta} and references therein for some examples of theoretically motivated models with extremely suppressed $\lambda_{H\Sigma}$).  

We will assume for simplicity that the scalar singlet field is  heavier that the Higgs boson, $m_\Sigma > 200\text{ GeV}$, and assume  $m_\Sigma \le \langle \Sigma \rangle$ in order to avoid non perturbative values of $\lambda_{\rm H\Sigma}$.

The DM density is generated by the decay of $\Sigma$  and is thus tied to the abundance of the latter, which in turn depends on the efficiency of the process $\Sigma\Sigma \leftrightarrow hh$ triggered by the portal like coupling $\lambda_{\rm H\Sigma}$ (by this we implicitly assume that, in case of very suppressed values of $\lambda_{\rm H\Sigma}$, the abundance of $\Sigma$ in the early stages of the evolution of the Universe is negligible). 
A proper description of the DM density requires the resolution of a system of coupled Boltzmann equations for the DM number density, as well as for the abundance of the $\Sigma$ field and possibly for the heavy pseudo-Dirac neutrinos, which also interact with $\Sigma$ -  also  including effects of entropy release. Further details of this computation can be found in the appendix. 

In the following we will present  analytical expressions which describe, to a good approximation, the  DM production mechanism. For simplicity we will assume that the pseudo-Dirac neutrinos are in thermal equilibrium (the case of non-equilibrium configurations substantially coincides with the studies already presented in~\cite{Petraki:2007gq,Merle:2013wta}) and with lifetimes such that the effects of entropy injection are not relevant.  As will be made clear, pseudo-Dirac neutrinos have a non trivial impact on DM production. We will thus for definiteness discuss two specific mass regimes, namely the case in which 
all the pseudo-Dirac neutrinos are lighter than $\Sigma$ and the case in which they  have instead similar or greater masses.

At high enough values of $\lambda_{H\Sigma}$,  the pair annihilation processes $\Sigma \Sigma \leftrightarrow hh$ maintain the field $\Sigma$ into thermal equilibrium.\footnote{Pair annihilation processes into fermion pairs are as well possible. For $m_\Sigma > m_h$, as assumed in this work, the relative rate is subdominant, being suppressed at least by a factor $v^2/m_\Sigma^2$.} Indeed,  
by comparing the $2 \rightarrow 2$ rate, associated to the thermally averaged cross-section $\langle \sigma v \rangle \sim 10^{-2}\times  \frac{\lambda_{\rm H\Sigma}^2}{m_\Sigma^2}$, with the Hubble expansion rate, the field $\Sigma$ can be considered to be  in thermal equilibrium in the Early Universe for $\lambda_{\rm H\Sigma} \ge \overline{\lambda}_{\rm H\Sigma}$ where:
\begin{equation}\label{lambdaFSIGMABAR}
 \overline{\lambda}_{\rm H\Sigma} \equiv 10^{-6} {\left(\frac{m_\Sigma}{100 \mbox{ GeV}}\right)}^{1/2}\ .
\end{equation}
On the contrary, its decay rate into DM is always suppressed compared to the Hubble rate due to the low value of the couplings $h_{\alpha\alpha}$ (see Eq.~(\ref{mumu})). The DM can thus be produced through the freeze-in mechanism from the decays of $\Sigma$ and its corresponding  abundance can be expressed as:
\begin{equation}
Y_{\rm DM}^{\rm FI}=\frac{135}{128 \pi^4} \sum_{I=5,8}\frac{|h_{\rm eff,I4}|^2}{g_{*}(T_{\rm prod})m_\Sigma}\left(1-\frac{m_I^2}{m_\Sigma^2}\right){\left(\frac{45 M_{\rm Pl}^2}{4 \pi^3 g_{*}}\right)}^{1/2}\ ,
\end{equation} 
where:
\begin{equation}
h_{\rm eff,I4}=\sum_{\alpha=1,3} U_{\rm I \alpha}^{T}\,h_{\alpha \alpha}\,U_{\alpha 4},
\end{equation}  
is an effective coupling taking into account all the decays $\Sigma \rightarrow N_I\, + \text{DM},\,\,\,I=4,8$ which are kinematically open.\footnote{Notice that since the scalar singlet field $\Sigma$ couples with all neutrinos, it can also decay into pseudo-Dirac states. However, the latter are in thermal equilibrium and thus no corresponding freeze-in production mechanism is possible.} The contribution to the relic DM density reads:
\begin{equation}
\Omega_{\rm DM}^{\rm FI} \approx 0.2 \sum_I {\left(\frac{|h_{\rm eff,I1}|}{10^{-8}}\right)}^2 \left(1-\frac{m_I^2}{m_\Sigma^2}\right) {\left(\frac{m_\Sigma}{200 \text{ GeV}}\right)}^{-1} \left(\frac{m_{s}}{1\text{ keV}}\right)\ . 
\end{equation}
On general grounds, out-of-equilibrium - i.e. after chemical decoupling - decays of $\Sigma$ may also contribute to DM production and 
the corresponding contribution 
to the DM density can be schematically expressed as:
\begin{equation}
\label{eq:ooe}
Y^{\rm SW}=B \ Y_\Sigma(T_{\rm f.o.})\,, \text{ where } B\equiv \sum_I b_I \text{Br}\left(\Sigma \rightarrow N_I N_1\right). 
\end{equation} 
In the above equation, $T_{\rm f.o.}$ is the standard freeze-out temperature of $\Sigma$ and  $b_I$ represents the number of DM particles produced per decay  for a given  decay channel. The branching ratio of the decay of $\Sigma$ into DM is given by:
\begin{equation}
\text{Br}\left(\Sigma \rightarrow N_I N_1\right)={{\sum_{I=1,5}^{}|h_{\rm eff,I1}|^2\left(1-\frac{m_I^2}{m_\Sigma^2}\right)}\over{\sum_{I,J=1,5}|h_{{\rm eff},IJ}|^2\left(1-\frac{{\left(m_I+m_J\right)}^2}{m_\Sigma^2}\right)+y_f^2 \sin^2 \alpha \left(1-\frac{4 m_f^2}{m_\Sigma^2}\right)}}\ ,
\end{equation}
where $\sin\alpha \propto \lambda_{H \Sigma}$ represents the mixing between $\Sigma$ and the Higgs boson.\footnote{This mixing exists only when the vev of the SM Higgs doublet is different from zero. Analogously to what we did for  the active-sterile mixing angle,  we have adopted in our computation a temperature dependent scaling function.} Due to the very low couplings  $h_{\alpha \alpha}$, the branching ratio of the  decay of $ \Sigma$ into DM is very suppressed  with respect to the branching ratio of the decay  into two fermions induced by the mixing with the Higgs boson, even for low values of the mixing itself. Furthermore,  the total lifetime of the scalar field is comparable with the freeze-out  timescale.
Consequently,  the out-of-equilibrium production is sizeable for $\lambda_{H\Sigma} \sim \overline{\lambda}_{H\Sigma}$ when the scalar field features an early decoupling, i.e. $x_{\rm f.o.}=m_\Sigma/T_{\rm f.o.}=1\ -\ 3$~\cite{Petraki:2007gq}.

Finally, the DM relic density can be estimated as:
\begin{align}
& \Omega^{\rm SW}_\text{DM} \approx 0.11 \left(\frac{m_\text{s}}{2 \text{ keV}}\right) \left(\frac{m_\Sigma}{1000 \text{ GeV}}\right) \left(\frac{B}{0.01}\right) {\left(\frac{\lambda_{\rm H\Sigma}}{10^{-6}}\right)}^{-2} \ .
\end{align}

In the case in which $\lambda_{\rm H\Sigma} \ll \overline{\lambda}_{\rm H\Sigma}$ (see Eq.~(\ref{lambdaFSIGMABAR})), the field $\Sigma$ is too feebly interacting to be in thermal equilibrium in the Early Universe. Assuming, for simplicity, a negligible abundance at early times, it can be nonetheless produced in sizeable quantities by freeze-in and then decay through out-of-equilibrium processes~\cite{Merle:2013wta}. The field $\Sigma$ is produced by the $2 \rightarrow 2$ processes mediated by the portal interactions as well as by the $2 \rightarrow 1$ processes, $N_I N_I \rightarrow \Sigma,\,\,I=5,8$ if the pseudo-Dirac neutrinos are lighter than $\Sigma$. \\The abundance of $\Sigma$ can be expressed as:
\begin{align}
\label{eq:SFI}
& Y_\Sigma^{\rm SFI}\approx \frac{M_{\rm Pl}}{1.66 g_{*}(T_{\rm prod})}\frac{1}{m_\Sigma} \left(\sum_{I,J} \frac{135}{128\pi^4} |h_{{\rm eff},IJ}|^2+\frac{45}{1024 \pi^6} \lambda_{\rm H\Sigma}^2\right) \ ,\\ \nonumber \text{  with}\\ \nonumber
& h_{{\rm eff,} IJ}=U^{T}_{I\alpha}h_{\alpha \alpha}U_{\alpha J}\ . 
\end{align} 
The two terms inside the parenthesis refer to the contributions from the  $2 \rightarrow 1$ (where the sum over $I,J$ runs over the pseudo-Dirac neutrinos in thermal equilibrium) and the $2 \rightarrow 2$ processes, respectively. The DM abundance is given, analogously to Eq.~(\ref{eq:ooe}), by $B \ Y_{\Sigma}^\text{SFI}$.

 In the case where the pseudo-Dirac neutrinos are heavier than $\Sigma$,  the DM generation process in the regime $\lambda_{H\Sigma} \geq \overline{\lambda}_{H\Sigma}$ proceeds along the same lines as described before. There is however the additional contribution from the decays $N_I \rightarrow h\, DM$, given by Eq.~(\ref{eq:heavy_fimp}) as well as a further freeze-in contribution from the decays $N_I \rightarrow \Sigma DM$ given by:
\begin{equation}
\label{eq:Sigma_FIMP}
\Omega^{\rm FI}_{\Sigma} \approx 2.16 \times 10^{-3} \sum_{I=5,8} g_I {\left(\frac{m_{\rm s}}{1 \text{ keV}}\right)} {\left(\frac{|h_{\rm eff, I\,4}|}{10^{-8}}\right)}^{2} {\left(\frac{m_{\rm I}}{1 \mbox{TeV}}\right)}^{-1}     \ .
\end{equation}

In the regime where  $\lambda_{H\Sigma} \leq \overline{\lambda}_{H\Sigma}$,  the $2 \rightarrow 1$ production channel for the field $\Sigma$ is replaced by the production from the decays of the pseudo-Dirac neutrinos through the processes, if kinematically open, $N_I \rightarrow N_J \Sigma,\,\,\,I=5,8,\,J=4,I-1$. In this scenario the abundance of $\Sigma$ reads:
\begin{equation}
\label{eq:SFI_bis}
Y_\Sigma^{\rm SFI}\approx \frac{M_{\rm Pl}}{1.66 g_{*}(T_{\rm prod})}\left[\frac{1}{m_\Sigma}\frac{45}{1024 \pi^6} \lambda_{\rm H\Sigma}^2+\sum_{I,J} \frac{135}{64\pi^4} \frac{|h_{{\rm eff},IJ}|^2}{m_I}\left(1-\frac{{\left(m_\Sigma+m_J\right)}^2}{m_I^2}\right)\right]\ .
\end{equation}

\begin{figure}[thb]
\begin{center}
\subfloat{\includegraphics[width=6.9 cm]{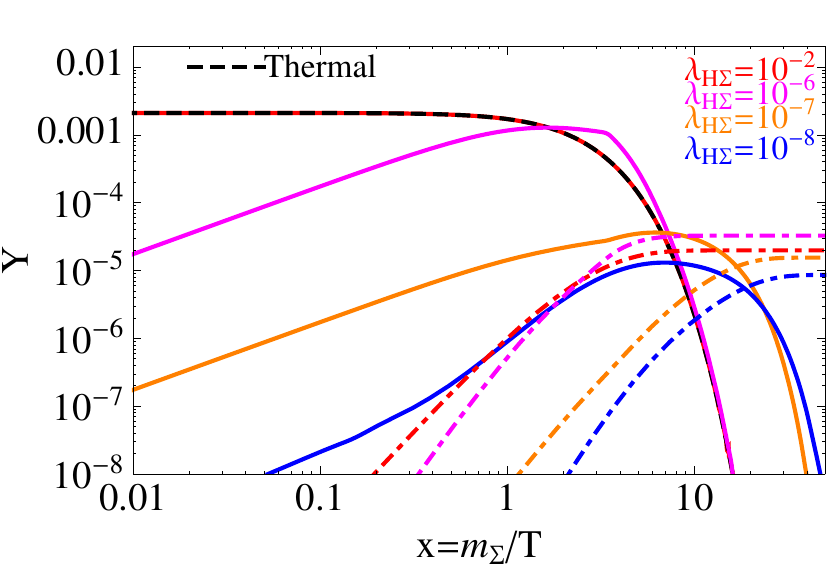}}\hspace*{0.5cm}
\subfloat{\includegraphics[width=6.9 cm]{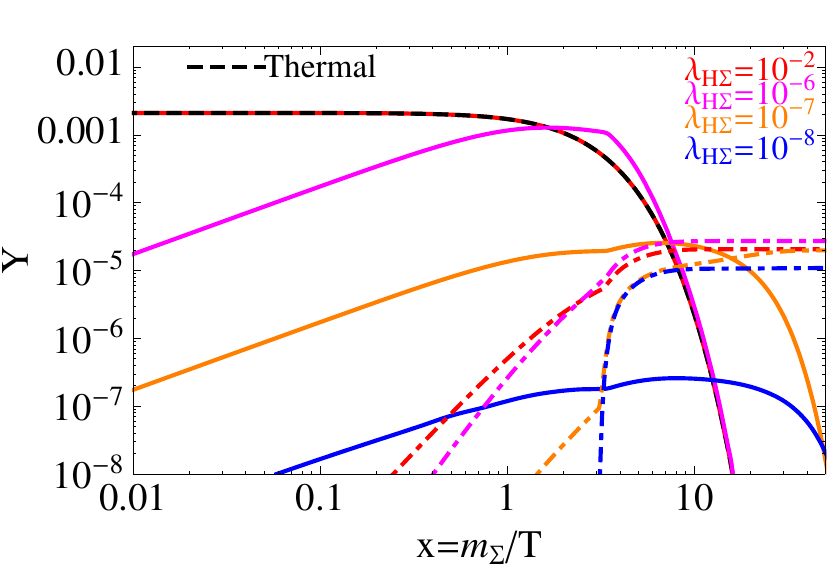}}
\caption{
Evolution of the abundance  of the $\Sigma$ field (solid lines) and of the DM (dot-dashed lines) for the four values of $\lambda_{H\Sigma}$ reported on the plot. The scalar field $\Sigma$ and the DM masses have been set to 500 GeV and 5 keV, respectively. The vev $\langle \Sigma \rangle$ has been fixed to 1 TeV. The masses of two the pseudo-Dirac pairs are, respectively 10 and 20 GeV (left panel) and 500 and 1000 GeV (right panel). In both cases, DM production through $N_I \rightarrow h + \text{DM}$ decays and the effects of entropy production are negligible.
 }
\label{fig:sterile_production_boltzmann}
\end{center}
\end{figure} 
 
The validity of the assumptions leading to the analytical approximations above has been confirmed by (and complemented) by numerically solving the Boltzmann equations for the system $\Sigma-$DM abundances.
We display in Figure~\ref{fig:sterile_production_boltzmann} the evolution of the abundances  of $\Sigma$ and of the DM, for a set of values of the the coupling $\lambda_{\rm H\Sigma}$ and for fixed values of $m_{\rm s}$, $m_\Sigma$ and $\langle \Sigma \rangle$ to $5 \text{ keV}$, $500 \mbox{ GeV}$  and $1 \mbox{ TeV}$, respectively.
 On the left panel the masses of the pseudo-Dirac pairs have been fixed to, respectively, 10 and 20 GeV, while on the right panel the chosen values are 500~GeV and 1~TeV. For this last case we have fixed the coupling $Y_\text{eff}$ of the pseudo-Dirac neutrinos with the Higgs boson and the DM-active neutrino mixing angle $\theta$ to, respectively, 0.01 and $10^{-6}$ in such a way that  the freeze-in production from the decays $N_I \rightarrow h +{\text{ DM}}$ gives a subdominant contribution, not exceeding 30$\%$. 

At higher values of $\lambda_{H\Sigma}$,  the abundance of $\Sigma$ traces its equilibrium value and the DM production occurs prevalently through the freeze-in mechanism. At lower values of $\lambda_{H\Sigma}$ the out-of-equilibrium production becomes important thus increasing the total DM relic density. 

For $\lambda_{H\Sigma} < \overline{\lambda}_{H\Sigma}$, the abundance of $\Sigma$ does not follow the equilibrium value but is an increasing function of time (as consequence of the freeze-in production from the $2 \rightarrow 2$ scatterings as well as from the $2 \rightarrow 1$ processes, for pseudo-Dirac neutrinos lighter than $\Sigma$, or from decays of the pseudo-Dirac neutrinos themselves in the opposite case) until its decay, which occurs at  later timescales compared to the case of high values of $\lambda_{H\Sigma}$.

\begin{figure}[thb]
\begin{center}
\subfloat{\includegraphics[width=6.0 cm]{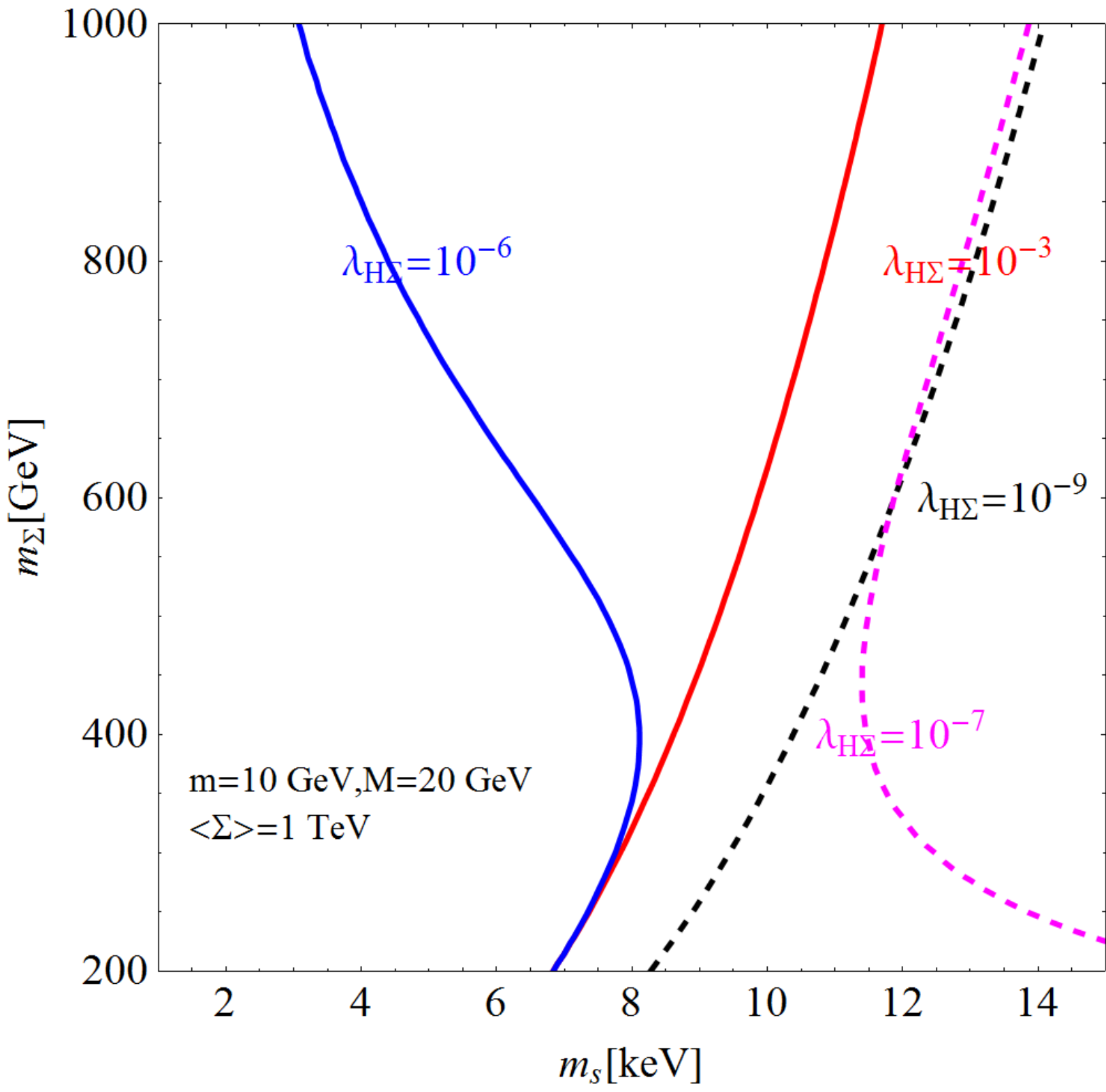}}
\hspace{5 mm}
\subfloat{\includegraphics[width=6.0 cm]{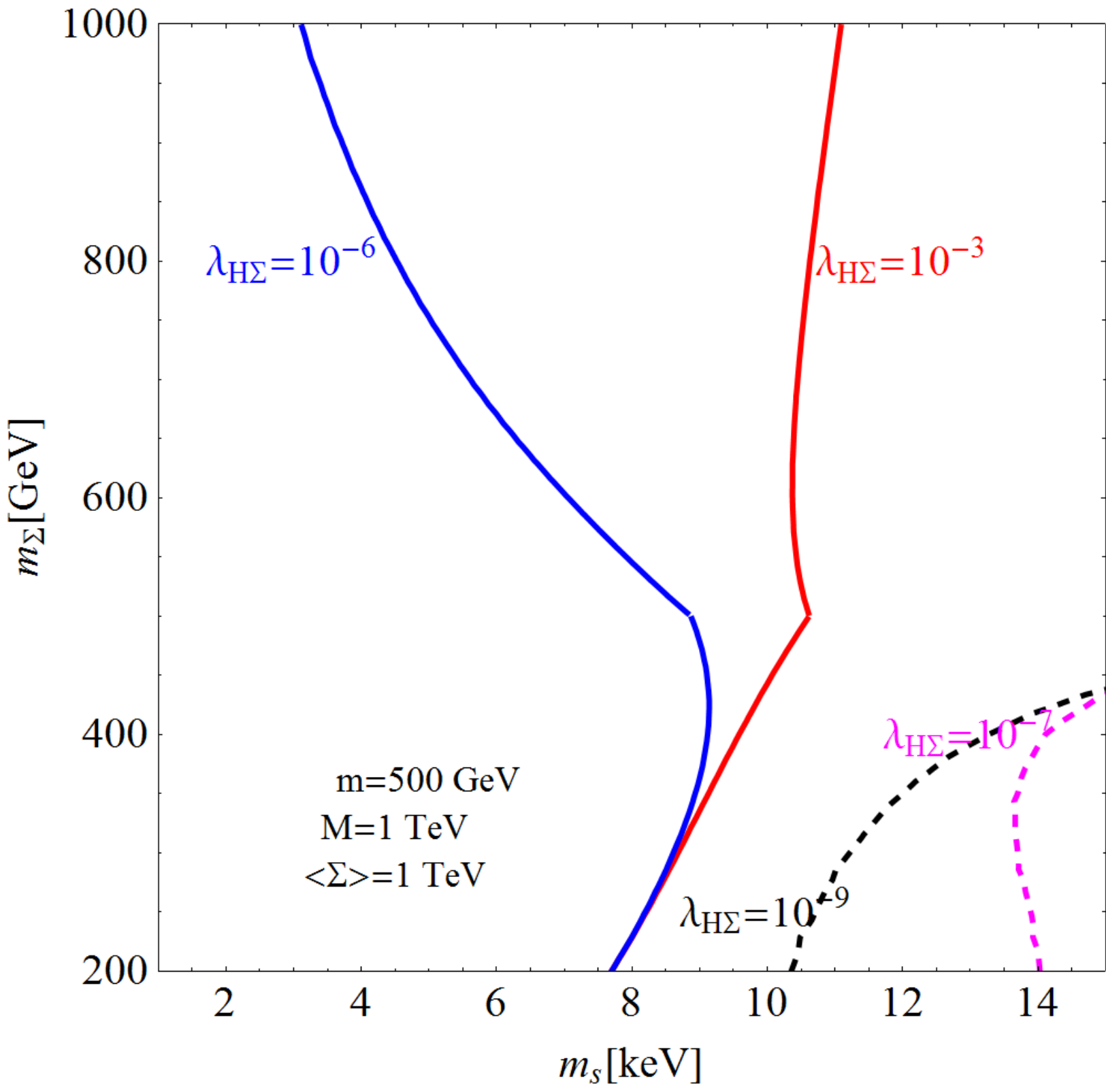}}
\caption{Contours of the cosmological value of the DM relic density in the plane $(m_s,m_\Sigma)$, for the values of $\lambda_{H\Sigma}$ reported in the plot. The other relevant parameters have been set as in Figure~\ref{fig:sterile_production_boltzmann}.
}
\label{fig:sterile_production_summary}
\end{center}
\end{figure} 

We report in Figure~\ref{fig:sterile_production_summary} the contours of the cosmological value of the DM relic density in the  $(m_s,m_{\rm \Sigma})$ plane for several  values of  the coupling $\lambda_{\rm H\Sigma}$ and for  the masses of the pseudo-Dirac neutrinos  considered in 
Figure~\ref{fig:sterile_production_boltzmann}. 
As already pointed out, for $\lambda_{H\Sigma}=10^{-3}$ the DM relic density is determined by the freeze-in mechanism and thus increases with the DM mass while decreasing with respect to $m_{\Sigma}$. For $\lambda_{H\Sigma}=10^{-6}$ the out-of-equilibrium production is instead the dominant contribution implying that $\Omega_s h^2 \propto m_s m_\Sigma$.
For $\lambda_{H\Sigma}=10^{-7}$ and $\lambda_{H\Sigma}=10^{-9}$, in the case of  light pseudo-Dirac neutrinos, the relic density is again proportional to the ratio $m_s/m_\Sigma$,  as expected from Eq.~(\ref{eq:SFI}).
 In the case of heavy pseudo-Dirac neutrinos, the regime $\lambda_{H\Sigma} \ll \overline{\lambda}_{H\Sigma}$ is substantially dominated by the freeze-in production of $\Sigma$ from the decays of   the pseudo-Dirac neutrinos and its subsequent out-of-equilibrium decays. The dependence on $m_\Sigma$ shown in Figure~\ref{fig:sterile_production_summary}  is due  to the kinematical factor in Eq.~(\ref{eq:SFI_bis}). 
We notice that the correct DM relic density, for the  chosen set of parameters, is achieved for DM masses between 1 - 15~keV. These results can be straightforwardly generalised in the case of entropy production from the pseudo-Dirac neutrinos. Indeed,  as can be seen from Figure~\ref{fig:sterile_production_boltzmann},  the DM production typically occurs at earlier stages compared to  the ones at which sizeable entropy production is expected (see previous section). As a consequence the instantaneous reheating approximation can be considered as valid and we can just rescale the DM relic density by a factor $S$. In this case,  the correct DM relic density is achieved for higher values of the DM masses. 

We emphasise, as already done in the previous section, that a complementary contribution to the DM relic density from DW production mechanism is in general present. The DM production related to the decays of $\Sigma$ allows to achieve the correct relic density without conflicting with the X-rays limits since it does not rely on the mixing with the active neutrinos;  this is not the case for the bounds from  structure formation.
However, applying  the limits on DM from structure formation is a very difficult task in our scenario since different DM production channels coexist, originating different dark matter distribution functions. A proper treatment would require to reformulate the bounds case by case by running suitable simulations, which lies beyond the scope of the present work. 
We will  nonetheless provide an approximate insight of how the latter bounds are altered, with respect to the conventional DW production mechanism, by taking  some representative examples.

In the following discussion, we consider  the case in which the DM is produced by the decays of the field $\Sigma$, either through freeze-in or through out-of-equilibrium decays.  An approximate reformulation of the limits from structure formation can be obtained by comparing  the average momentum of  DM at the keV scale with the one corresponding to DW production and by rescaling the lower limit on the DM mass with  the shift between these two quantities.  
The DM distribution function in the various cases of production from decay has been determined in e.g.~\cite{Petraki:2007gq} and~\cite{decay}). 
The dark matter produced through freeze-in is typically generated at temperatures of the order of the mass of $\Sigma$. Its average momentum depends only on the temperature and can be simply expressed, at temperatures of the order of the keV, as~\cite{Shaposhnikov:2006xi}:
\begin{equation}
\left.\left(\frac{\langle p \rangle}{T}\right)\right |_{T \sim \text{keV}}  \simeq 0.76  \ S^{-1/3}\ ,
\end{equation}
sensitively lower than the corresponding result  (of $\sim 2.83$) in the case of DW production. A similar result holds as well in the case of DM produced  through freeze-in from  the decays of the pseudo-Dirac neutrinos.\footnote{The DM distribution function can be obtained by solving the same Boltzmann equation as in~\cite{Petraki:2007gq} and  by replacing the Bose-Einstein distribution for the decaying state with a Fermi-Dirac function. 
The difference in the final result is of order one.} 
In the case in which the DM is prevalently produced out-of-equilibrium, the timescale of production varies with the lifetime of $\Sigma$ and the distribution function tends to be warmer as the latter increases.
\begin{figure}[thb]
\begin{center}
\includegraphics[width=8.7 cm]{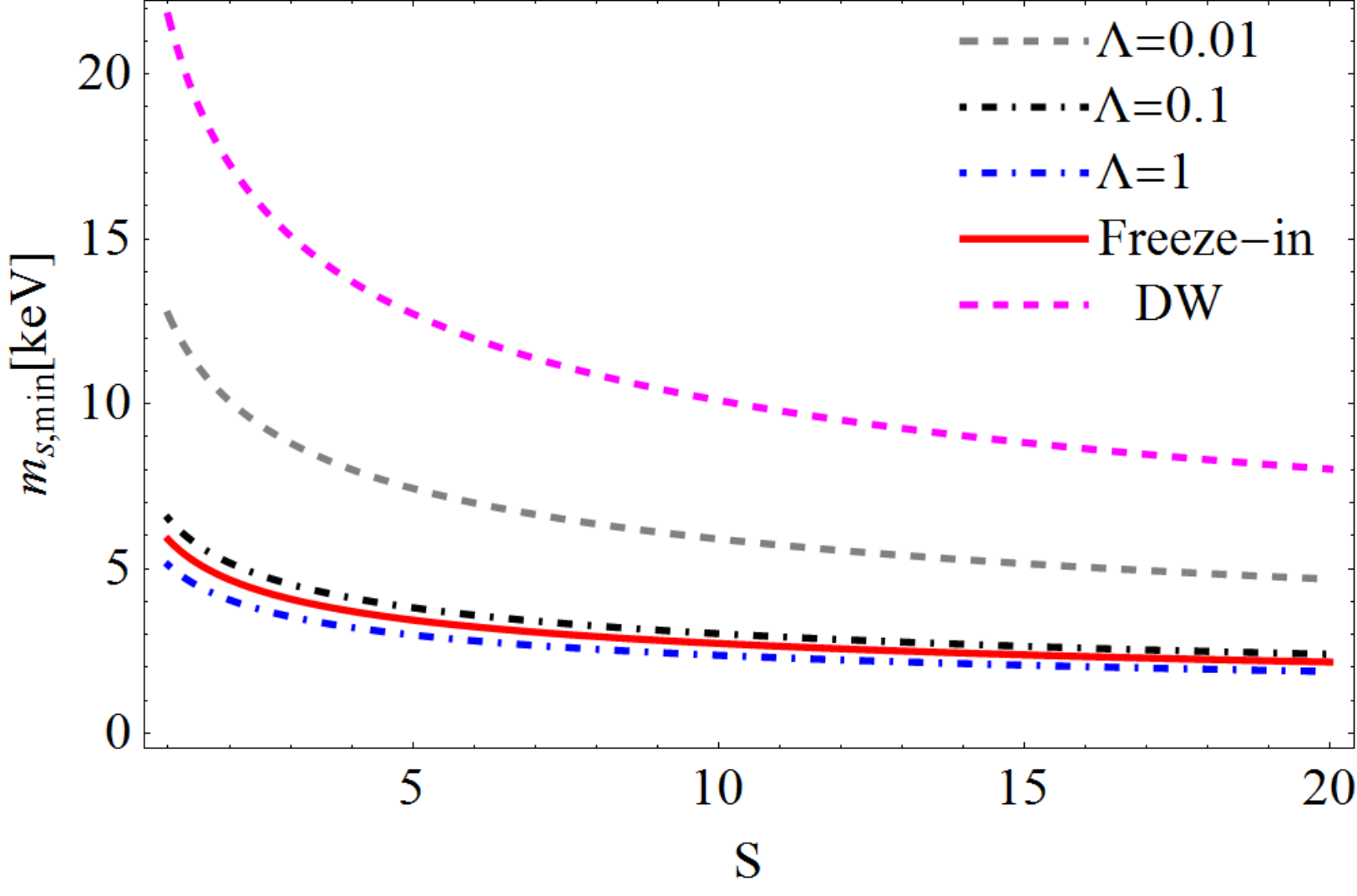}
\caption{Lower limit on the DM mass from Lyman-$\alpha$ 
 as a function of the entropy dilution factor $S$. The limit refers to the cases of dominant freeze-in production from the decay of the scalar field $\Sigma$ and dominant production from out-of-equilibrium decays  for the parameter $\Lambda= 0.01,\ 0.1,\   1$ (defined in Eq.~(\ref{def:Lambda})). We also report for comparison the   corresponding limit in the case of dominant DW production mechanism.
} 
\label{fig:S_Lyalpha}
\end{center}
\end{figure}

We report in Figure~\ref{fig:S_Lyalpha} the lower limit\footnote{Since we are here assuming that the lightest sterile neutrino is the only DM component, we adopt the most updated limit. This actually refers to a thermal  relic density. It can be reformulated in term of a limit on non-resonant DW production by using the formula given in~\cite{Viel:2005qj}.} from Lyman-$\alpha$ on the DM mass, obtained by applying our approximate rescaling to the limit presented in~\cite{Viel:2013fqw}, for some scenarios of DM production mechanism, namely freeze-in and out-of-equilibrium production for different decay rates of $\Sigma$ parametrised through the dimensionless quantity:
\begin{align}\label{def:Lambda}
& \Lambda=\frac{5 \overline{h}^2}{8 \pi m_\Sigma}{\left(\frac{45 M_{\rm Pl}^2}{4 \pi^3 g_{*}}\right)}^{1/2}\ , \nonumber\\
\text{where,}\nonumber\\
& \overline{h}^2=\sum_{I,J} |h_{{\rm eff},IJ}|^2 \left(1-\frac{{\left(m_I+m_J\right)}^2}{m_\Sigma^2}\right)+y_f^2 \sin\alpha^2 \left(1-\frac{4 m_f^2}{m_\Sigma^2}\right)\ . 
\end{align} 
As can be noticed,  in the most favourable cases, namely freeze-in or out-of-equilibrium production with $\Lambda \geq 1$ (corresponding to $m_\Sigma \lesssim 600\text{ GeV})$, the limit from Lyman-$\alpha$ is relaxed to approximately 5~keV, in absence of entropy injection, and to further low values for the case $S>1$. We remark again that these results assume that all the DM is produced by the decay of $\Sigma$ (into heavy neutrinos).

Interestingly, in the case in which  a sizeable contribution from DW production mechanism is also  allowed,  the DM distribution would feature a Warm and a ``colder'' component and thus the "(2,3) ISS" model  could potentially realise a mixed Cold + Warm DM scenario. This would constitute an  intriguing solution to some tensions with observation from structure formation (see e.g.~\cite{Weinberg:2013aya} and references therein). 
This possibility should be thoroughly investigated by means of   numerical simulations since the analytical estimates presented above are not valid for multi-component distributions. This will be thus left for a dedicated study.      

\section{Conclusion}\label{Sec: Conclusions}
In this study we have considered the possibility of simultaneously addressing the dark matter problem and the neutrino mass generation mechanism. We have focused on the truly minimal inverse seesaw  realisation - the "(2,3) ISS" model-  fulfilling all phenomenological and cosmological requirements and  which provides a Warm Dark Matter candidate (for a mass of the lightest sterile state around the keV). 

We have conducted a comprehensive analysis taking into account the several possibilities of neutrino mass spectra. In most of the parameter space the DM can be produced only through active-sterile transitions according to the DW production mechanism, accounting, in the most favourable case, for at most  $\sim43\%$ of the relic  DM abundance, without conflict with observational constraints.
 This situation can be improved  for two specific choices of the spectrum of the heavy pseudo-Dirac neutrinos. Firstly, one can consider the  case of moderately light, i.e. $\sim 1\ -\ 10$ GeV, pseudo-Dirac neutrinos. These states can dominate the energy density of the Universe and produce entropy at the moment of their decay, altering the impact of DM on structure formation. However the constraints from dark matter indirect detection  are still too severe and the allowed DM fraction is increased only up to $\sim 50\%$. The second possibility relies upon relatively heavy, $\sim 130\text{ GeV}-1\text{ TeV}$, pseudo-Dirac pairs, which can produce the correct amount of DM through their decays. In this kind of setup it is also possible for the "(2,3) ISS"   to account for the recently reported $3.5$ keV line in galaxy cluster spectra. 
In the final part of this work, we have proposed a minimal extension of the "(2,3) ISS" with the addition of a scalar singlet (at the origin of the lepton number violating masses of the sterile fields) which allows to achieve the correct DM relic density for generic values of the masses of pseudo-Dirac neutrinos. The latter  can still participate, at various levels, to the DM production. 

\acknowledgments
We are deeply grateful to J. Lesgourgues, O. Ruchayskiy and M. Viel for their valuable and important comments.
We are thankful to Y. Mambrini  for enlightening discussions. We also thank L. Covi and M. Garny for their interesting remarks and discussions. 
We acknowledge  support from the European Union FP7 ITN
INVISIBLES (Marie Curie Actions, PITN-\-GA-\-2011-\-289442). G.A. thanks the LPT Orsay and the Cern Theory Division for the warm hospitality
during part of the completion of this work.

\appendix 

\section{Boltzmann equation for sterile neutrinos produced from decay}

In this appendix we briefly describe the numerical treatment used to validate and complement the results presented in the main text. On general grounds one should solve a system of coupled Boltzmann equations for the abundance of the $\Sigma$ field as well as all the 5 extra neutrinos of the ISS scenario. As already
mentioned we will focus on the case in which the pseudo-Dirac neutrinos can be regarded in thermal equilibrium during the DM production phase. 
This allows to focus on a system of two coupled Boltzmann equations whose general form is:
\bee
\label{eq:num_sys}
 \frac{d n_{\Sigma}}{dt}+3 H n_{\Sigma}&=&-B \langle \Gamma \rangle n_{\Sigma} -(1-B) \langle \Gamma \rangle \left(n_{\Sigma}-n_{\Sigma, \rm eq}\right)\nonumber\\
&&+ \sum_I \tilde{B}_I \langle \Gamma_{N_I} \rangle n_{I}+\sum_I \left(1-\tilde{B}_I\right) \langle \Gamma_{N_I} \rangle \left(n_I-n_{I,\rm eq}\right) \nonumber\\ 
&&-\langle \sigma v \rangle \left(n_\Sigma^2-n_{\Sigma, \rm eq}^2\right), \nonumber\\
& \nonumber\\
 \frac{d n_{\rm DM}}{dt}+ 3 H n_{\rm DM}&=& B \langle \Gamma \rangle n_{\Sigma}\nonumber\\
&& +\sum_I \tilde{B}_I \langle \Gamma_{N_I} \rangle n_{I,\rm eq}+\sum_I \langle \Gamma\left(N_I \rightarrow h\, +\text{ DM}\right)\rangle n_{I,\rm eq} \nonumber\\
&& +DW.
\eee
The first equation traces the time evolution of the field $\Sigma$. The first row on the right-hand side represent the decay of $\Sigma$, respectively into at least one DM particle and only into pseudo-Dirac neutrinos, if kinematically allowed. Since the latter are assumed in thermal equilibrium this second term is balanced by a term accounting for inverse decays and thus vanishes if $\Sigma$ is in thermal equilibrium. On the contrary the first term is not balanced by an inverse decay term since the DM has too weak interactions to be in thermal equilibrium and then can be assumed to have a negligible abundance at early stages; this originates the freeze-in production channel. The second row represents the decays, if kinematical allowed, of the pseudo-Dirac neutrinos into $\Sigma$ and another neutrino. The factor $\left(n_I-n_{I,\rm eq}\right)$ assumes that $\Sigma$ is in thermal equilibrium and disappears if the pseudo-Dirac neutrinos are as well in thermal equilibrium. Here we have again distinguished the decay term into DM, which is non balanced by the inverse process, and the decay term into final thermal states (this distinction holds only if $\Sigma$ is in thermal equilibrium. In the regime $\lambda_{\rm H\Sigma} \ll \overline{\lambda}_{\rm H\Sigma}$ the second row should be replaced by the term $\sum_I \langle \Gamma_{N_I} \rangle n_{I}$). The last term finally represents the annihilation processes of $\Sigma$. $B$ and $\tilde{B}_I$ represent the effective branching fractions of decay of, respectively, $\Sigma$ and the pseudo-Dirac neutrinos. $\langle \Gamma \rangle$ and $\langle \sigma v \rangle$ represent the conventional definitions of the thermal averages~\cite{Gondolo:1990dk}:
\bee
 \langle \Gamma \rangle &=& \Gamma \frac{K_1(x)}{K_2(x)},\nonumber\\
 \langle \sigma v \rangle &=& \frac{1}{8 m_\Sigma^4 T K_2^2(m_\Sigma/T)} \int_{4 m_\Sigma^2}^{\infty} ds \sigma_{\rm ann} \left(s-4 m_\Sigma^2\right) \sqrt{s} K_1\left(\sqrt{s}/T\right),\,\,\,\,\,\,\sigma_{\rm ann} \propto \frac{\lambda_{H\Sigma}^2}{s} \nonumber \\
&=& \frac{\lambda_{H\Sigma}^2}{4 m_\Sigma^2 x^2}F(x),
\eee
where the function $F(x)$ is determined by numerically solving the integral above.

The second equation traces the DM number density. The first two rows represent the DM production from, respectively, $\Sigma$ and the pseudo-Dirac neutrinos. The term labelled $DW$ represents instead the contribution associated to production from oscillation processes. In the parameter space of interest the two production processes, decay and oscillations, occur at well separated time scales; as a consequence we can drop the DW term from the equations and possibly add its contribution to the final relic density.

In order to account possible effects of entropy injection from the decays of the pseudo-Dirac neutrinos the system above should be completed with a third equation accounting for the non conservation of the entropy (see e.g.~\cite{Arcadi:2011ev}). On the other hand it has been shown that the pseudo-Dirac neutrinos can dominate the energy budget of the Universe and inject sizeable amount of entropy only  at very late times, compared to the DM production from decay which occurs at temperature close to the mass scale of $\Sigma$ (a possible exception is the case $\lambda_{\rm H\Sigma} \ll \overline{\lambda}_{\rm H\Sigma}$). To a  good approximation we can thus stick on a system of the form~(\ref{eq:num_sys}) and apply a posteriori possible entropy effects.  

For simplicity we will describe two specific examples, namely all the pseudo-Dirac neutrinos lighter or heavier than $\Sigma$. In the first case all the source terms associated to the decays of the pseudo-Dirac neutrinos can be dropped. Moving to the quantities $Y_{\Sigma, \rm DM}=n_{\Sigma,\rm DM}/s$ and $x=m_\Sigma/T$ as, respectively, dependent and independent variables, the system reduces to:
\bee
\frac{dY_\Sigma}{dx} &=& -\frac{1}{16 \pi}\frac{\tilde{h}^2 m_\Sigma}{H x}\frac{K_1(x)}{K_2(x)} \left(Y_\Sigma -(1-B) Y_{\Sigma,eq}\right)\nonumber\\
&&-\frac{45 \lambda_{\rm H\Sigma}^2 m_\Sigma}{512 \pi^7 g_{*} H x^6}F(x)\left(Y^2_{\Sigma}-Y^2_{\Sigma,eq}\right),\nonumber\\
 \frac{dY_{\rm DM}}{dx} &=& \frac{1}{16 \pi}\frac{\tilde{h}^2 m_{\Sigma}}{H x}\frac{K_1(x)}{K_2(x)} B Y_\Sigma, \nonumber\\
\tilde{h}^2 &=& \sum_{IJ} |h_{{\rm eff},IJ}|^2 \left(1-\frac{{\left(m_I+m_J\right)}^2}{m_\Sigma^2}\right),\,\,\,\,\nonumber\\
 B &=& \frac{\sum_I |h_{\rm eff,I1}|^2\left(1-\frac{{\left(m_I\right)}^2}{m_{\Sigma}^2}\right)}{\sum_{I,J}|h_{\rm eff,I1}|^2\left(1-\frac{{\left(m_I+m_J\right)}^2}{m_{\Sigma}^2}\right)+y_f^2 \sin^2 \alpha \left(1-\frac{4 m_f^2}{m_\Sigma^2}\right)},
\eee
where $Y_{\Sigma,\rm eq}=\frac{45}{4 g_{*}\pi^4}x^2 K_2(x)$ and $H$ is the Hubble expansion rate $H=\sqrt{\frac{4 \pi^3 g_{*}}{45}}\frac{m_\Sigma^2}{x^2 M_{\rm Pl}}$. This last expression assumes that during the phase of DM generation the Universe is radiation dominated, this is reasonable since we have shown in the main text that the number density of the heavy neutrinos tends to dominate at  low temperatures.

The numerical solution of this system has been presented in the left panel of Figure~\ref{fig:sterile_production_boltzmann} for some sample values of the relevant parameters.

The analytic expressions provided in the text correspond instead to suitable limits in which this set of equations can be solved analytically.  In the regime $\lambda_{{\rm H}\Sigma} \geq \overline{\lambda}_{{\rm H}\Sigma}$ the right-hand side of the equation for the DM is dominated, at early times, by the annihilation term and we have simply $Y_{\Sigma}=Y_{\Sigma, \rm eq}$. In this regime we have only to solve the equation of the DM substituting $Y_{\Sigma, \rm eq}$ on the right-hand side. The equation can be straightforwardly integrated for:
\begin{equation}
Y_{\rm DM}= \frac{45}{1.66\,64 \pi^5 g_{*}^{3/2}}\frac{M_{\rm Pl}}{m_{\Sigma}}\sum_I |h_{\rm eff,I1}|^2\left(1-\frac{{\left(m_I\right)}^2}{m_{\Sigma}^2}\right) \int x^3 K_1(x) dx.
\end{equation}     
For late enough decays we can integrate the Bessel function from zero to infinity thus obtaining the freeze-in contribution to DM relic density:
\begin{equation}
Y^{\rm FI}_{\rm DM}= \frac{135}{1.66\,128 \pi^4 g_{*}^{3/2}}\frac{\sum_I |h_{\rm eff,I1}|^2 M_{\rm Pl}}{m_{\Sigma}},
\end{equation}
where we have neglected, for simplicity, the kinematical factors in this last expression.      
 At late times the only relevant terms in the equation are the decay terms, and the DM equation can be again integrated with initial condition $Y_{\Sigma}=Y_{\Sigma, \rm eq}(x_{f.o.})$, obtaining the SuperWimp contribution to the DM relic density.
 In the regime $\lambda_{\rm H\Sigma} < \overline{\lambda}_{\rm H\Sigma}$ instead the abundance of the $\Sigma$ field is always below the equilibrium value. We can thus drop the term proportional to $Y_{\Sigma}$ in the first Boltzmann equation which can be directly integrated over $x$. Assuming again enough late decays we can carry the integration until infinity obtaining Eq.~(\ref{eq:SFI}).   

In the case in which the pseudo-Dirac neutrinos are heavier than $\Sigma$ the system of Boltzmann equations is modified as:
\bee
\frac{dY_\Sigma}{dx} &=& -\frac{1}{16 \pi}\frac{\tilde{h}^2 m_\Sigma}{H x}\frac{K_1(x)}{K_2(x)} Y_{\Sigma}\nonumber\\
&& +\frac{1}{16 \pi}\sum_I \tilde{B}_I \frac{\overline{h}_I^2 m_I}{H x}\frac{K_1(x)}{K_2(x)}Y_{I} +\frac{1}{16 \pi}\sum_I \left(1-\tilde{B}_I\right) \frac{\overline{h}_I^2 m_I}{H x}\frac{K_1(x)}{K_2(x)}\left(Y_I-Y_{I,eq}\right)\nonumber\\
&&+\frac{45 \lambda_{\rm H\Sigma}^2 m_\Sigma}{512 \pi^7 g_{*} H x^6}F(x)\left(Y^2_{\Sigma}-Y^2_{\Sigma,eq}\right), \nonumber\\
 \frac{dY_{\rm DM}}{dx}&=&\frac{1}{16 \pi}\frac{\tilde{h}^2 m_{\Sigma}}{H x}\frac{K_1(x)}{K_2(x)} B Y_\Sigma \nonumber\\
&& +\frac{1}{16 \pi}\sum_I \tilde{B}_I \frac{\overline{h}_I^2 m_I}{H x}\frac{K_1(x)}{K_2(x)}Y_{I}+\frac{1}{16 \pi}\sum_I \tilde{B}_I \frac{Y^2_{\rm eff}\sin^2\theta m_I}{H x}\frac{K_1(x)}{K_2(x)}Y_{I},\nonumber\\
& \nonumber\\
\overline{h}_I^2 &=& \sum_{J} |h_{{\rm eff},IJ}|^2 \left(1-\frac{{\left(m_\Sigma+m_J\right)}^2}{m_I^2}\right),\non 
\tilde{B}_I &=& \frac{|h_{\rm eff,I4}|^2\left(1-\frac{{\left(m_\Sigma\right)}^2}{m_{I}^2}\right)}{\sum_{J}|h_{{\rm eff},IJ}|^2\left(1-\frac{{\left(m_\Sigma+m_J\right)}^2}{m_{I}^2}\right)}.
\eee
In the regime $\lambda_{\rm H\Sigma} > \overline{\lambda}_{\rm H\Sigma}$ the second row of the equation for $Y_\Sigma$ can be neglected and we can fix again $Y_{\Sigma}=Y_{\Sigma,\rm eq}$ and derive analytical solutions for the DM relic density through analogous steps as above. In the case $\lambda_{\rm H\Sigma} \ll \overline{\lambda}_{\rm H\Sigma}$ we have to replace the second row of the equation for $Y_\Sigma$ with $\frac{1}{16 \pi}\sum_I\frac{\overline{h}_I^2 m_I}{H x}\frac{K_1(x)}{K_2(x)}Y_{I},\,\,\,\,Y_I=Y_{I,\rm eq}$ and we can again fix $Y_\Sigma=0$ on the right-hand side. Analytical solutions are reliable if the timescales of production and decay of $\Sigma$ are well separated, otherwise one should refer to the numerical treatment.

\bibliographystyle{hieeetr}

\end{document}